\documentclass[fleqn,usenatbib]{mnras}

\usepackage{newtxtext,newtxmath}
\usepackage[T1]{fontenc}
\usepackage{ae,aecompl}
\usepackage{CJKutf8}
\usepackage{graphicx}
\usepackage{amsmath}
\usepackage{natbib}
\usepackage[monochrome]{color}
\usepackage{xcolor}
\usepackage{multirow}

\title[HIR4: neutral hydrogen sky simulation]{HIR4: cosmology from a simulated neutral hydrogen full sky using Horizon Run 4}
\author[Asorey et al]{\parbox{\textwidth}{ Jacobo Asorey$^1$\thanks{E-mail: Jacobo.Asorey@ciemat.es}, David Parkinson$^1$, Feng Shi$^1$, Yong-Seon Song$^1$, Kyungjin Ahn$^2$,  Juhan Kim$^3$, Jian Yao$^4$, Le Zhang$^{4,5}$ and Shifan Zuo$^{6,7}$} \\ \\
$^1$ Korea Astronomy and Space Science Institute, Yuseong-gu, Daedeok-daero 776, Daejeon 34055, Korea\\
$^2$ Department of Earth Sciences, Chosun University, Gwangju 61452, Korea\\
$^3$ Center for Advanced Computation, Korea Institute for Advanced Study, 85 Heogiro, Dongdaemun-gu, Seoul 02455, Korea\\
$^4$ School of Physics and Astronomy, Shanghai Jiao Tong University, Shanghai 200240, P. R.  China\\
$^5$ School of Physics and Astronomy, Sun Yat-Sen University, 2 Daxue Road, Tangjia, Zhuhai, 519082, P.R. China \\
$^6$ Key Laboratory of Computational Astrophysics, National Astronomical Observatories, Chinese Academy of Sciences,\\
 Beijing 100101, P. R. China\\
$^7$ Department of Astronomy and Tsinghua Center for Astrophysics, Tsinghua University, Beijing 100084, P. R. China}

\date{Accepted 2020 April 21. Received April 20; in original form 2020 January 10}

\pubyear{2020}
\volume{495}
\pagerange{1788-1806}

\begin{document}
\label{firstpage}
\maketitle

\begin{abstract}
The distribution of cosmological neutral hydrogen  will provide a new window into the large-scale structure of the Universe with the next generation of radio telescopes and surveys. The observation of this material, through  21cm line emission, will be confused by foreground emission in the same frequencies. Even after these foregrounds are removed, the reconstructed map may not exactly match the original cosmological signal, which will introduce systematic errors and offset into the measured correlations. In this paper, we simulate future surveys of neutral hydrogen using the Horizon Run 4 (HR4) cosmological N-body simulation. We generate HI intensity maps from the HR4 halo catalogue, and combine with foreground radio emission maps from the Global Sky Model, to create accurate simulations over the entire sky. We simulate the HI sky for the frequency range 700-800 MHz, matching the sensitivity of the Tianlai pathfinder. We test the accuracy of the fastICA, PCA and log-polynomial fitting foreground removal methods to recover the input cosmological angular power spectrum and measure the parameters. We show the effect of survey noise levels and beam sizes on the recovered the cosmological constraints. We find that while the reconstruction removes power from the cosmological 21cm distribution on large-scales, we can correct for this and recover the input parameters in the noise-free case. However, the effect of noise and beam size of the Tianlai pathfinder prevents accurate recovery of the cosmological parameters when using only intensity mapping information.
\end{abstract}


\section{Introduction}

The distribution of matter on large-scales  has  provided an important cosmological probe, allowing for measurement of the cosmological parameters by probing both the initial conditions that generated the seeds of structure, and also the physics that causes the structures to grow and develop. This data has come from large area extragalactic surveys, mainly targeting galaxies in the optical, which act as tracers of the underlying dark matter density fluctuations. While the first surveys were initially at very low-redshifts (e.g. CfA survey, \citep{GELLER897}, 2df \citep{2001MNRAS.328.1039C}), their reach has increased as the technology has improved. Results from these surveys are independent from, but complimentary to, that of the more distant cosmic microwave background (CMB).

These surveys  have an advantage over the CMB, as the density distribution of the Universe  can be measured from galaxies in three dimensions, rather than the two-dimensional surface from which the CMB photons are emitted. This property allows for an increase in the sampling, as a particular scale can be measured in more directions, leading to a decrease in sample variance (e.g. \cite{2007PhRvD..76h3004S}). It also allows for types of measurements impossible in two-dimensions, for example the radial components of the BAO giving a direct probe of the Hubble rate $H(z)$ \citep{2003ApJ...594..665B,2003PhRvD..68f3004H,2003ApJ...598..720S}. However, there are disadvantages as well, since the structures at late time will have undergone some non-linear evolution under gravity which requires more complex modelling. There is also the greater  observational time and effort required by galaxy redshift survey, as opposed to a simpler photometric survey \citep{2005MNRAS.363.1329B}.  

Most of these surveys of the matter distribution have so far been in the optical and near-infrared, as the combination of technological lead and targets available in the this wavelength range has made these sources most accessible.  However, these surveys will soon reach a natural limit, as the expansion of the Universe will redshift the spectra of these objects out of the observable range, leading to a `redshift desert' in the range $1.4 < z < 3$.  The next generation of radio observatories offers an alternative in this region, through the measurement of 21cm emission of neutral atomic hydrogen (HI). This line emission should be observable at greater redshifts, as radio telescopes can span a much larger range of frequencies than the optical wavelength band. This signal can be used either in the same manner as traditional optical galaxy spectroscopy, to identify the redshift of individual galaxies (HI galaxy surveys) \citep{2018ApJ...861...49H}, or by measuring the total HI intensity in a larger sky area (pixel), known as HI intensity mapping \citep{Bharadwaj2001}. Intensity Mapping in particular can be carried out much faster than optical spectroscopic surveys, giving the opportunity to conduct full-sky sky surveys and so measure the matter distribution on the largest scales \citep{PhysRevLett.111.171302}.

However, since the spontaneous hydrogen spin-flip transition is highly forbidden, with a long mean lifetime, and the intergalactic density of cold neutral hydrogen is low, the 21cm signal is relatively weak ($T \sim 1mK$). This weak signal can be overwhelmed by foreground radio emission at the same observed frequency ($T \sim 1K$ or greater at around 800 MHz). These foreground radio emissions  include the thermal radiation from the ionosphere of the Earth, the galactic synchrotron, free-free emission from ionized regions both galactic and extra-galactic, and radio point sources contaminants. Separating the cosmological HI signal from the non-HI foreground will be difficult, and a number of different methods have been proposed.  All of these methods assume a smooth, large-scale frequency-dependence of the foregrounds, that can be modelled and removed for each pixel, leaving the small-scale fluctuations that correspond to the 21cm density fluctuation. These include: fastICA which removes the foregrounds using independent component analysis in frequency space \citep{2012MNRAS.423.2518C,FastICA}, PCA (principal component analysis) and ICA (independent component analysis) \citep{2015MNRAS.447..400A}, the Correlated Component Analysis (CCA) method \citep{10.1093/mnras/stu2601}, Karhunen-Loeve Decomposition \citep{Shaw_2014,PhysRevD.91.083514}, Generalized Morphological Component Analysis (GMCA) \citep{2013MNRAS.429..165C}. 

The most important questions are then  how well the cosmological parameters can be measured by future data sets (precision), and how much bias can be introduced into the measurements by the foreground removal process (accuracy). In order to settle both of these questions, forecasts need to move beyond the assumption of Gaussianity, and simple power spectrum distributions of the fluctuations, and consider accurate sky simulations, based on large-scale N-body simulations. In this paper we address both questions, with particular emphasis on accuracy, showing how biases or offsets in the posterior probability distribution of the cosmological parameters (offsets relative to the input parameter values of the simulations) are introduced by the effect of removing the foreground contamination, and can be corrected for using simulations. 

Another method to increase the accuracy of the recovered cosmological parameters from the 21cm radio sky would be to cross-correlate the intensity map with a galaxy catalogue that covers the same area of the sky and redshift range. Since the optical galaxy sample would not have the same systematic errors and uncertainties introduced by the radio foreground removal process, the cross-power spectrum should be a more accurate representation of the underlying density field. This would also be straightforward to demonstrate with simulations, if the halos can be populated by a galaxy distribution that matches the planned survey. In a forthcoming paper we will simulate this cross-correlation using the same prescription and analysis approach as we have here \citep{Feng2020}.

For the HIR4 (HI with Horizon Run 4) project, we have created a full simulation and analysis pipeline. Some alternative approaches in the literature consider hydrodynamical simulations \citep{2018ApJ...866..135V} or fast simulations based on lognormal density field realizations  \citep{2014MNRAS.444.3183A}. However, in our case, we have started with the dark matter-only particle  Horizon Run 4 N-body simulation, and populated the halos with clouds of neutral hydrogen.  When using N-Body simulations, there has to be a necessary compromise between detailed hydrodynamical simulations and fast simulations in terms of volume and resolution. But the mass limit in large volume N-body simulations does not allow us to access all the hydrogen, as some will be located in halos with masses below the limit. This can be solved by scaling the simulated maps with observational measurements of the neutral hydrogen density, $\Omega_{\mathrm{HI}}$. Having simulated the cosmological signal, we applied foreground radio emission at the appropriate wavelengths based on the best available current data. In addition, we included the estimated instrumental noise and array beaming effect for the Tianlai pathfinder \citep{2011SSPMA..41.1358C,2018SPIE10708E..36D} in our simulations. We then masked the sky and applied reconstruction techniques to remove the foregrounds and reconstruct the cosmological signal. Finally, we measured the angular power spectra of the 21cm temperature maps, and measured the cosmological parameters from these. Since we have complete control over every step of the process, any mis-match or offset between the final cosmological results and the initial cosmological parameters set in the simulation will then provide a test of the steps we have taken and the assumptions we have made in the analysis process.

In section \ref{sec:method} we describe the methodology we use to generate our simulations and analyse our data, including details about the Horizon Run 4 simulations,  generating the simulated the cosmological and foreground sky, and reconstructing the cosmological signal, and measuring the angular power spectra. In section \ref{sec:results} we show the sky maps that we generate, the measured angular power spectra for different cases, and the cosmological constraints. We summarise the forecast precision that these measurements will have in terms of the linear bias and growth rate of structure, and also address the accuracy at which the different reconstruction methods recover the input cosmology. In section \ref{sec:discussion} we summarise our findings, and make recommendations for future analysis of real data.

\section{Methodology}
\label{sec:method}

\subsection{Cosmology theory}

The intensity of the 21cm brightness temperature field $T_b$, as a function of spatial position ($\vec{x}$) and cosmological time $t$, can be considered as a perturbation relative to the homogeneous mean temperature $\bar{T}_{b}$ evaluated at time $t$, such that
\begin{equation}
    \label{eqn:temp_pert}
    \Delta T_{b}(\vec{x},t) = T_{b}(\vec{x},t)-\bar{T}_{b}(t)\,.
\end{equation}
If we assume that the statistics of the density of neutral hydrogen track the statistics of the overall matter density, then we can make the usual assumption that the two are related through some bias parameter $b$, and rearrange Eq. \ref{eqn:temp_pert} to give the 21cm field temperature in terms of the matter density perturbation $\delta$,
\begin{equation}
    T_{b}(\vec{x},t) = \bar{T}_{b}(t) [ 1 + b(\vec{x},t)\delta(\vec{x},t) ] \,.
\end{equation}
If we now transform from a real space position and homogeneous time coordinate to a measured sky direction $\vec{n}$ and redshift $z$, we need to include the effects due to redshift-space distortions, giving
\begin{equation}
    T_{b}(\vec{n},z) = \bar{T}_{b}(z) \left[ 1 + b(\vec{x},t)\delta(\vec{x},t) +\frac{1+z}{H(z)}n^i\partial_i(\vec{n}\cdot\vec{v}) \right] \,.
\end{equation}

The clustering of fluctuations is described by the anisotropic power spectrum in Fourier space that can therefore be written (assuming the Kaiser formula for redshift-space distortions \citep{1987MNRAS.227....1K} and linear perturbation theory) as
\begin{equation}
    P_{\rm 21cm}(k,\mu,z) = \bar{T}_{b}(z)^2P_{\delta\delta}(k,z)\left[b_{\rm HI}(z)+f(z)\mu^2\right]^2\,,
\end{equation}
where $\mu=\cos{\theta_{los}}$ is the ratio between a given mode and the radial modes given by the line of sight.
We describe the linear growth rate $f(z)$ theoretically by parameterizing it with a growth index $\gamma$ \citep{2007APh....28..481L},
\begin{equation}
    f(z) = \Omega_m(z)^\gamma\,,
\end{equation}
and as we are assuming a $\Lambda$CDM model with Einstein gravity, the growth index is given by $\gamma=0.545 \  $\citep{1984ApJ...284..439P,1991MNRAS.251..128L}.

\subsection{Generating Sky Maps}

\label{sec:sky_maps}
We have created maps of 21cm emission for different configurations.  We start from an N-body simulation and then we proceed to create neutral hydrogen mass catalogues, that are finally converted in brightness temperature maps. We also add foregrounds and receiver noise to the combination.

\subsubsection{Horizon Run 4 simulations}
\label{HR4sims}
Since we are interested in simulating wide field 21cm intensity mapping surveys, we start by using a halo catalogue from a very large volume N-body simulation as the initial framework for the neutral hydrogen. We have used the Horizon Run 4 \citep{2015JKAS...48..213K} simulations to create our intensity mapping maps. 
It is an N-body simulation run on a box of $L_{box}=3150h^{-1}$ Mpc. From the dark matter particles in the light-cone a halo catalogue was built using a Friend-of-friends algorithm. The minimum halo mass is $2\cdot 10^{11}h^{-1}M_{\odot}$ which correspond to 25 dark matter particles in the original lightcone. It is based on a flat $\Lambda$CDM cosmology with a matter density of $\Omega_m=0.26$, a Hubble parameter at redshift zero of $H_0=72~{\rm kms}^{-1}{\rm Mpc}^{-1}$, and an amplitude of fluctuations on scales of $8~ h^{-1}{\rm Mpc}$ of $\sigma_8 = 1/1.26$.

\begin{figure}
\centering
\includegraphics[trim = 0cm 0cm 0cm 0cm, width=0.45\textwidth]{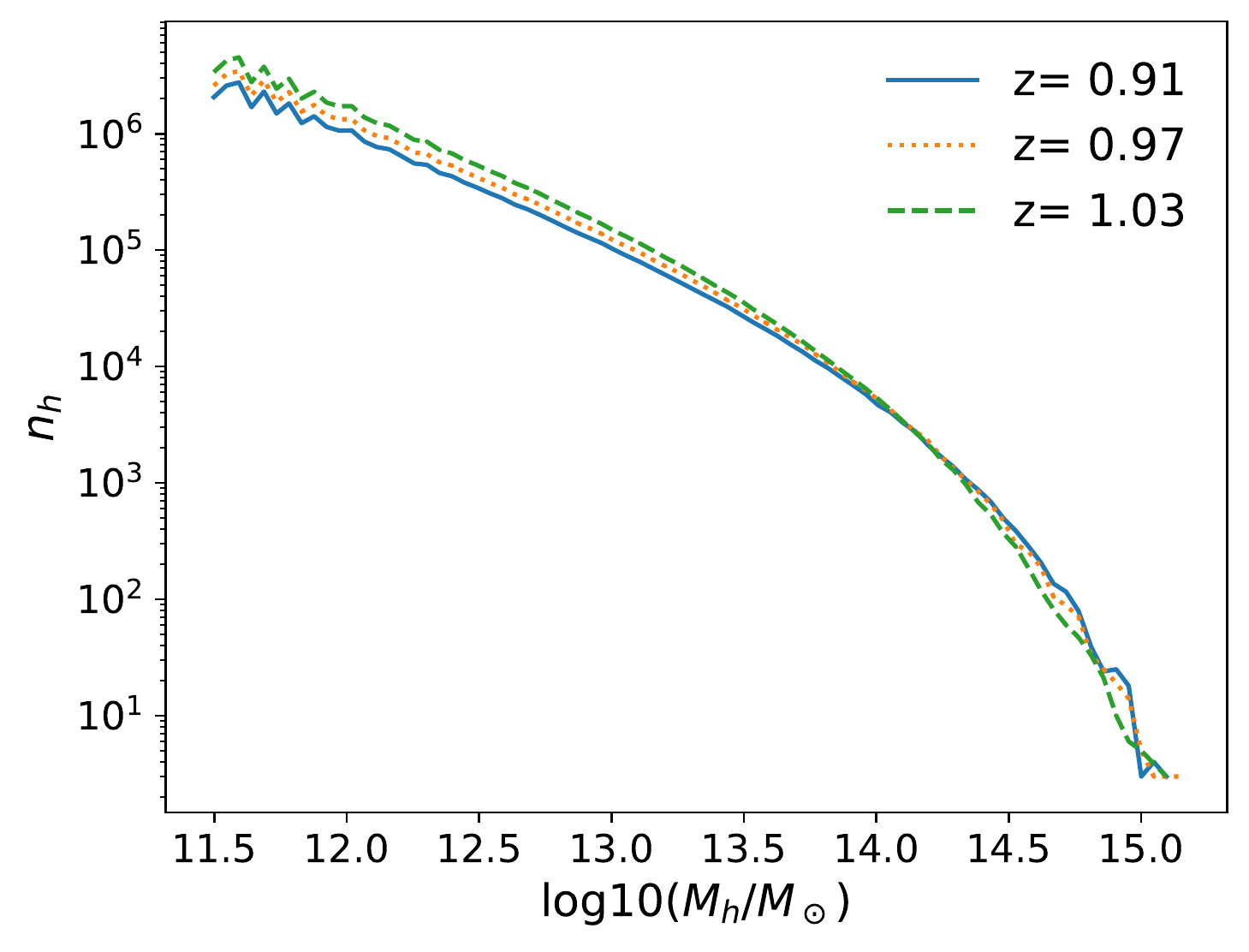}
\caption{Distribution of halo masses for halos selected in the expected redshift range for the Tianlai pathfinder. The sharp cut-off for the small masses is due to the resolution limit of the HR4 simulation.}
\label{fig:figNH}
\end{figure}

The total number of halos in the catalogue is $1,654,566,127$. In Fig. \ref{fig:figNH}, we show the mass halo function for halos in the redshift range sampled by Tianlai and the range we consider in this paper. The simulation also include the velocities of the haloes. Using this velocities, we can create a catalogue that includes the linear redshift distortions. This redshift $z_v$ is given by the combination of the cosmological redshift and the peculiar velocity one \citep{2016ApJ...832..103L,2019arXiv190712639D}
\begin{equation}
    (1+z_{v}) = (1+z_{\rm true})(1+v_z/c),
\end{equation}
where $z_{\rm true}$ is the true redshift given by the Hubble-Lema\^itre flow and $z_v$ is the redshift that includes the effect of the peculiar velocity in the radial direction $v_z$.

\subsubsection{Neutral Hydrogen mass modelling}
As we are focusing in the post-reionization Universe, we can assume almost all the neutral hydrogen content around redshift $z=1$ is inside dark matter haloes \citep{2018ApJ...866..135V,2020MNRAS.tmp..569S}.
We assign masses of neutral hydrogen to dark matter halos based on the halo mass $M_h$ and virial velocity $v_c$, following the halo model for neutral hydrogen developed in successive improvements in \cite{2015MNRAS.454..218B,2016MNRAS.458..781P,2017MNRAS.464.4008P}. There are other approaches (e.g. \citet{2019JCAP...09..024M}), but since we expect our signal to be dominated by the most massive neutral hydrogen halos, we find this prescription to be sufficient for our use.

\begin{figure}
\centering
\includegraphics[trim = 0cm 0cm 0cm 0cm, width=0.45\textwidth]{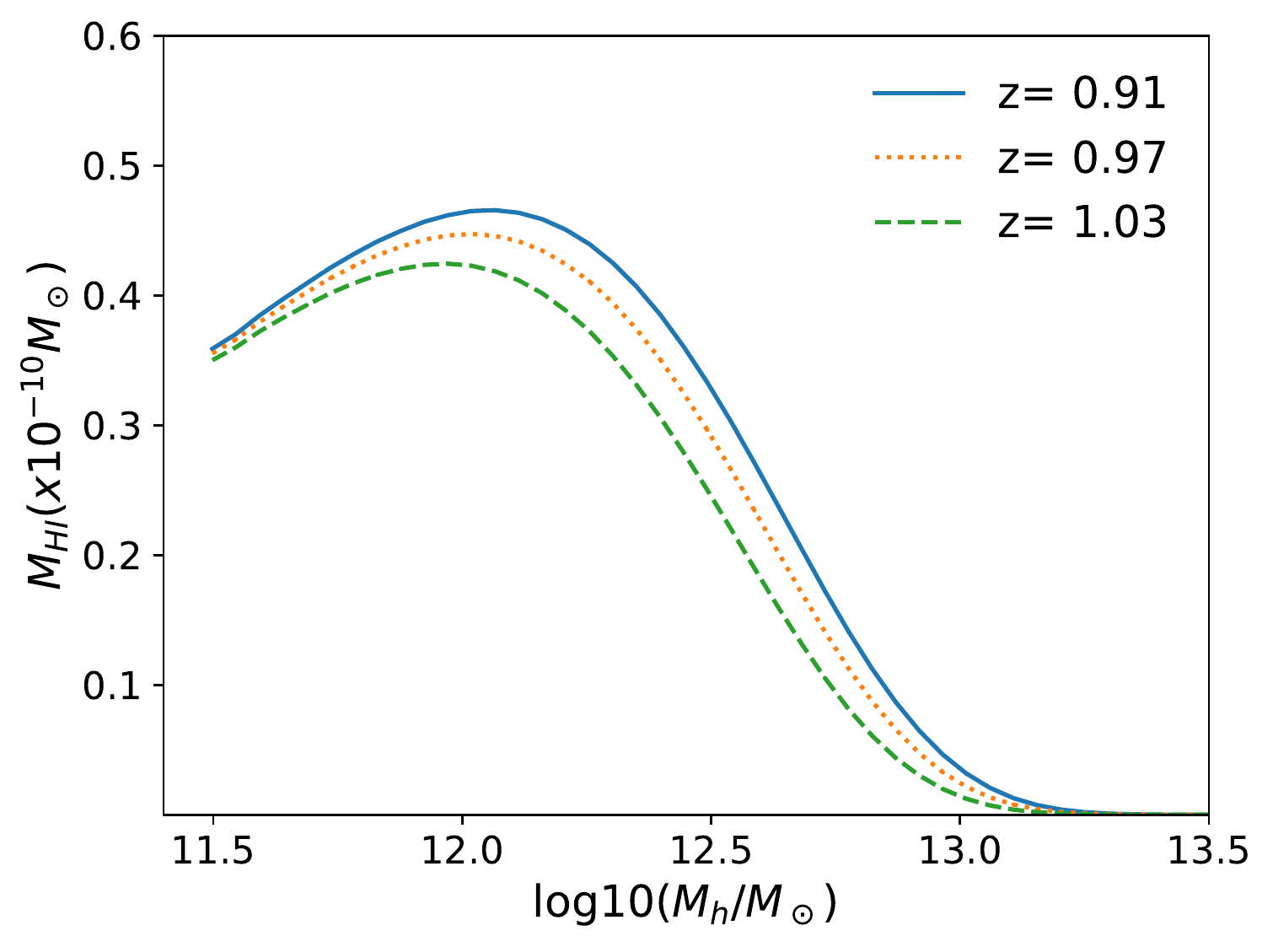}
\caption{Distribution of hydrogen masses following Eq. \ref{eq:mhi_halo_model} at three redshifts within the Tianlai redshift range and the HR4 simulation.}
\label{fig:figNH_mass}
\end{figure}

In \cite{2017MNRAS.464.4008P} the  mass of neutral hydrogen hosted in a dark matter halo of mass $M_h$ is given by:
\begin{eqnarray}
M_{\mathrm{HI}}(M_h)=&&f_{\mathrm{HI}} f_{H,c}M_h\left(\frac{M_h}{10^{11}h^{-1}M_\odot}\right)^\beta \nonumber\\
&&\exp \left[-\left(\frac{v_{c}^{\rm min}}{v_c(M_h)} \right)^3\right]\exp\left[-\left(\frac{v_c(M_h)}{v_{c}^{\rm max}} \right)^3\right],
\label{eq:mhi_halo_model}
\end{eqnarray}
where $f_{\mathrm{HI}}$ is a multiplicative constant that corresponds to the amount of neutral Hydrogen with respect to the fraction of cosmic Hydrogen, $f_{H,c}=(1-Y_p)\Omega_{b0}/\Omega_{m0}$ where $Y_p=0.24$ is the primordial Helium abundance. The model includes a logarithmic slope $\beta$ and two velocity cut-offs $v_{c}^{\rm min}$ and $v_{c}^{\rm max}$. The reason for the velocity cut-offs is that low-mass halos are not capable of keeping the neutral hydrogen while massive halos heat the gas and it stops being neutral. The values used for our simulation, which are partially based on \cite{2017MNRAS.464.4008P} best fit to data, are $f_{\mathrm{HI}}=0.17$, $\beta=-0.55$, ${v_{c}^{\rm min}}=30$ km/s and ${v_{c}^{\rm max}}=200$ km/s. We have used different values for the cut-off velocity parameters in order to find similar values of the hydrogen bias, as given in observations. In order to define the virial velocity, we have used the spherical collapse model as in 
\begin{equation}
    v_c(M_h) = \sqrt{\frac{G_N M_h}{R_c}},
\end{equation}
where $R_c(M_h)$ is the virial radius. Following \cite{2016MNRAS.458..781P}, we determine the virial radius by
\begin{equation}
    R_c(M_h) = 46.1 \hbox{kpc} \left(\frac{\Delta_v \Omega_m h^2}{24.4} \right)^{-1/3}\left(\frac{1+z}{3.3} \right)^{-1}\left(\frac{M_h}{10^{11}M_\odot} \right)^{1/3}.
\end{equation}
We use $\Delta_v$ given by the solution to a spherical top hat perturbation collapse for a virialized halo for the flat $\Lambda$CDM Universe, $\Omega_k=0$, \citep{1980lssu.bookP,1996MNRAS.282263E,1998ApJ...495...80B} where 
\begin{equation}
    \Delta_v = 18\pi^2+82x -39x^2,
\end{equation}
and $x=\Omega_m(z)-1$. We show in Fig. \ref{fig:figNH_mass}, the average neutral hydrogen mass for a given halo mass at three different redshifts. We can observe that the mass distribution decreases with the massive halos cut-off while we do not reach the cutt-off in the least massive halos as the resolution of the Horizon Run 4 is not enough to reach the quenching scale. 

\begin{figure}
\centering
\includegraphics[trim = 0cm 0cm 0cm 0cm, width=0.45\textwidth]{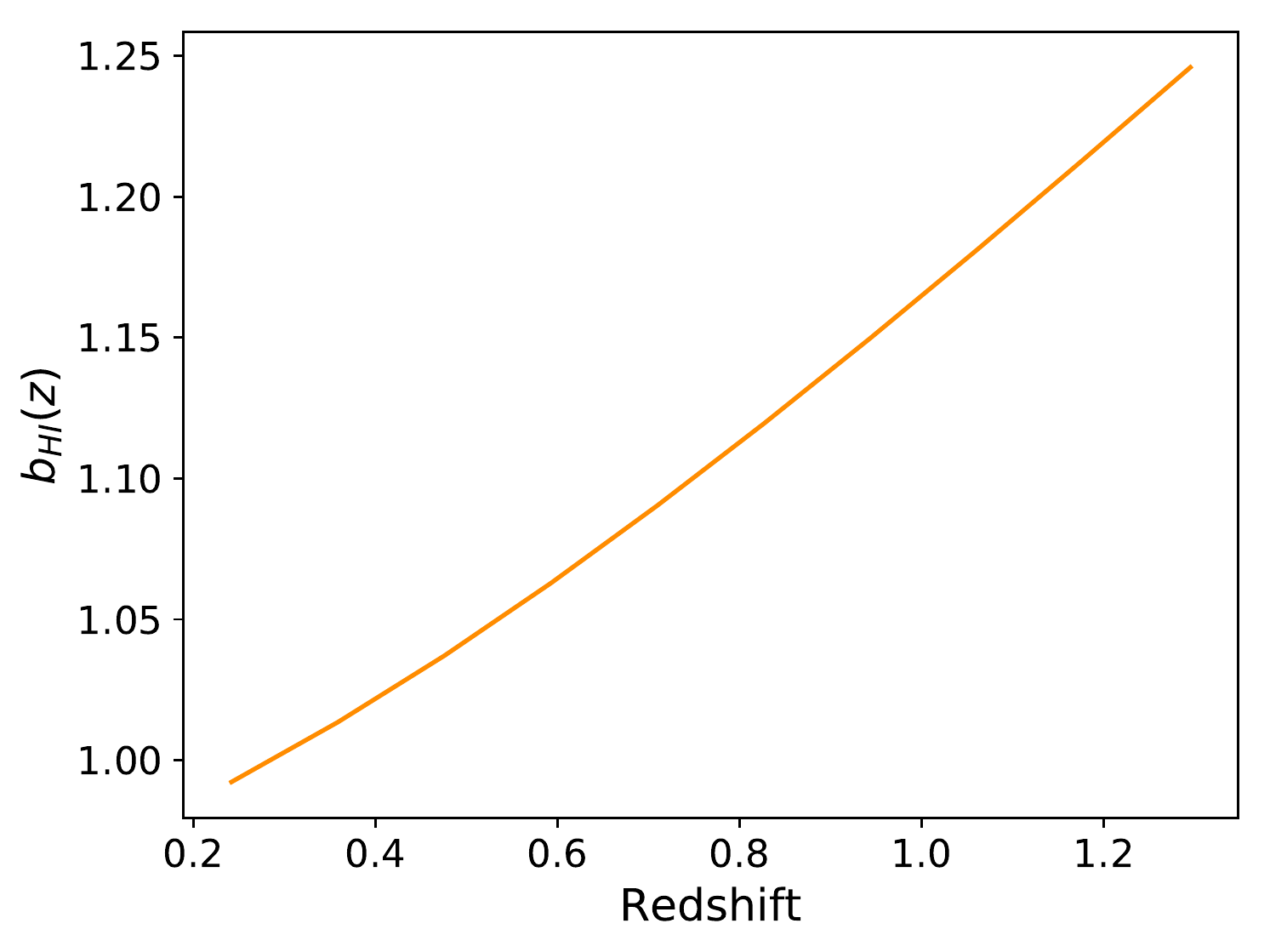}
\caption{Bias of neutral hydrogen for the halos in our sample for the the full redshift range of the Horizon Run simulation. The bias have been measured using Eq. \ref{eq:hi_bias} and we measured the halo mass function and the neutral hydrogen mass distribution from the simulation.}
\label{fig:figbhi}
\end{figure}

We can define the bias of neutral hydrogen, $b_{\mathrm{HI}}(z)$ as
\begin{equation}
b_{\mathrm{HI}}(z)=\frac{\int{dMn(M,z)M_{\mathrm{HI}}(M,z)b(M,z)}}{\int{dMn(M,z)M_{\mathrm{HI}}(M,z)}},   
\label{eq:hi_bias}
\end{equation}
while the neutral hydrogen density parameter is:
\begin{equation}
\Omega_{\mathrm{HI}}(z)=\frac{\rho_{\mathrm{HI}}}{\rho_{c,0}}=\frac{1}{\rho_{c,0}}\int_0^{\infty}{dMn(M,z)M_{\mathrm{HI}}(M,z)}.    
\label{eq:omegahi}
\end{equation}
A more in-depth discussion of the neutral hydrogen bias, using hydrodynamical simulations, can be found in \citet{2019MNRAS.484.5389A,Wang2019}.
In Fig. \ref{fig:figbhi} we show the bias estimated using Eq. \ref{eq:hi_bias}. The values are consistent with previous studies in the literature \citep{2010ApJ...718..972M}. We use this parameter as a benchmark parameter to test our simulated catalogues.

%

\subsubsection{Brightness temperature maps}
Once we have assigned hydrogen masses to the halos in our simulation, we can continue to the next step, the creation of intensity maps. In order to do so, we define a redshift bins configuration and a given pixelization resolution. Then, we stack all the hydrogen masses for all the haloes in each cube defined by an angular pixel and a redshift bin. The corresponding mass $M_{\mathrm{HI}}$ is what we use to create the temperature maps. Following \cite{2013MNRAS.434.1239B,2015ApJ...803...21B} we define the 21cm brightness temperature as:
\begin{equation}
T_{b}({\bf \hat{n}},z)=\frac{3h_{\mathrm{Pl}}c^3A_{12}}{32\pi k_bm_h\nu_{21}^2}\frac{(1+z)^2}{H(z)}\rho_{\mathrm{HI}}({\bf \hat{n}},z)
\label{eq:t_21},
\end{equation}
where $k_b$ is the Boltzmann constant, $h_{\mathrm{Pl}}$ is the Planck constant, $m_h$ is the mass of the neutral hydrogen atom, $A_{12}$ is the quantum efficiency, $c$ is the speed of light, and $\rho_{\mathrm{HI}}$ is the density of neutral hydrogen in the volume given by the frequency and area, $d\nu$ and $d\Omega$, which correspond the  frequency (redshift) and pixel bins, respectively.

One caveat of our method is that we cannot access all the halo masses that host neutral hydrogen as the HR4 simulation has a lower limit for the mass of the haloes. As we do not have access to all the halo masses, we do not completely sample the full $\rho_{\mathrm{HI}}$ in a given volume cell. This lack of mass will produce a smaller brightness amplitude $T_{b}$ than the expected one in nature. As the 21 cm cosmological signal has a low amplitude, reconstruction from a foreground dominated map becomes more difficult as the amplitude of the cosmological signal from neutral hydrogen becomes smaller. Therefore, we need to take into account the shortfall of neutral hydrogen in the simulation by scaling the average brightness temperature according to observations.

From Eq. \ref{eq:omegahi}, we see that $\rho_{\mathrm{HI}}$ is proportional to the density parameter of neutral hydrogen, $\Omega_{\mathrm{HI}}$. We can use measurements of this parameter in order to calibrate the mean temperature of our 21cm maps. We have decided to follow the definition given in \cite{2018arXiv181102743S,2019arXiv190401479C}. The approach is based on a polynomial fit to the $\Omega_{\mathrm{HI}}$ data compiled in \cite{2015MNRAS.452..217C}.

Both analysis \citep{2018arXiv181102743S,2019arXiv190401479C} define this fit as:
\begin{equation}
    \Omega_{\mathrm{HI}}(z) = 0.00048 + 0.00039z - 0.000065z^2.
    \label{eq:omegahiska}
\end{equation}

\begin{figure}
\centering
\includegraphics[trim = 0cm 0cm 0cm 0cm, width=0.45\textwidth]{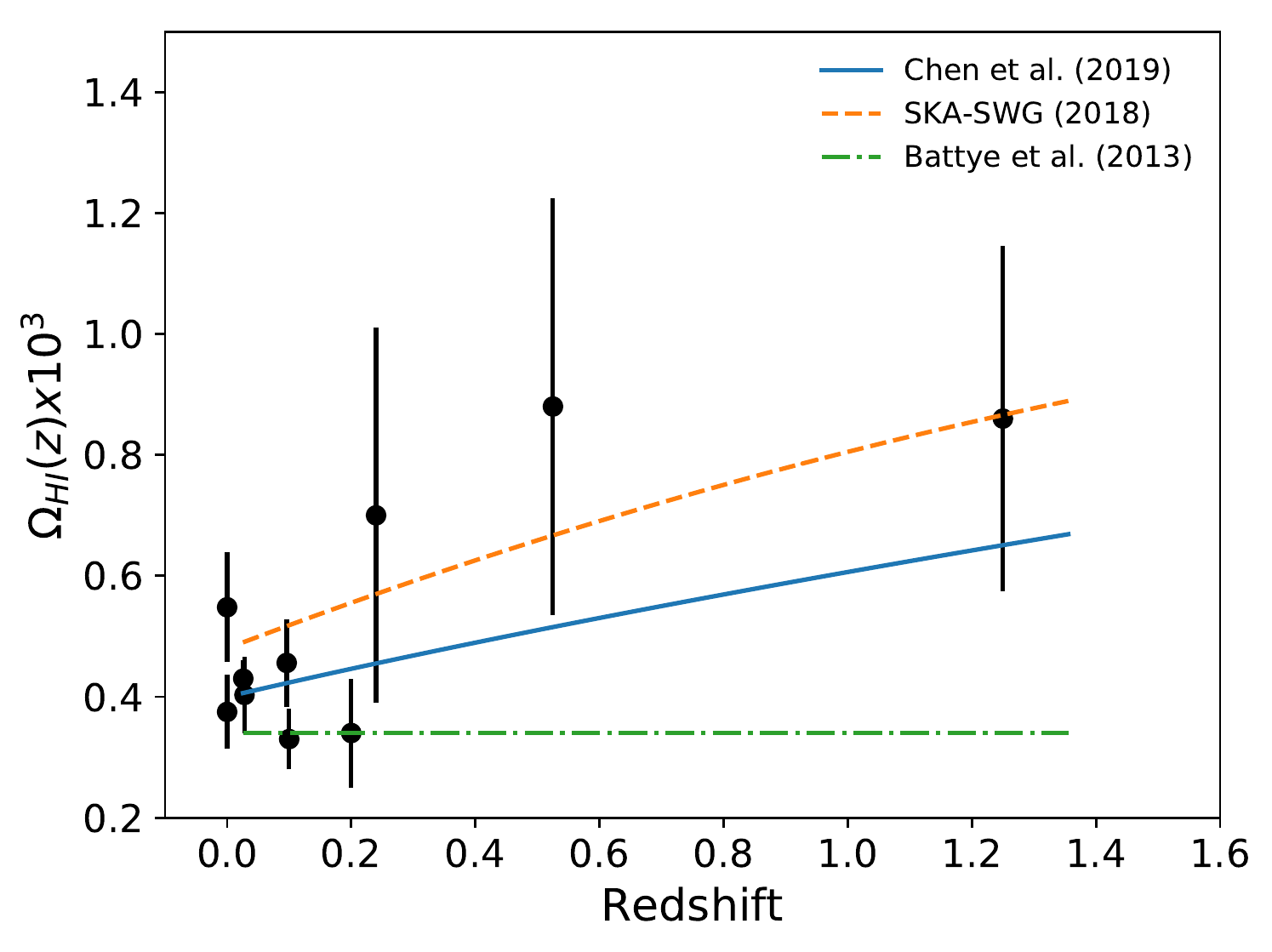}
\caption{Comparison between theoretical descriptions of $\Omega_{\mathrm{HI}}$ given by lines and measurements compiled in \citet{2015MNRAS.452..217C} (black circles).}
\label{fig:omhi_mean_freq}
\end{figure}
We can compare this approach with other models in the literature. Using the same data compilation \cite{2015MNRAS.452..217C} but a different model for the redshift dependency given by $\Omega_{\mathrm{HI}}\propto(1+z)^{0.6}$.  In \cite{2013MNRAS.434.1239B}, the density parameter is assumed constant and fitted to low redshift data where $\Omega_{\mathrm{HI}}=2.45e-4$. We can see the comparison in Fig. \ref{fig:omhi_mean_freq}.

Then, we re-scale the brightness temperature $\bar{T}_{b}^{\mathrm{old}}$ give by Eq. \ref{eq:t_21} in each map and pixel
\begin{equation}
     T_{b}^{\mathrm{new}}({\bf \hat{n}},z) = \frac{T_{b}^{\mathrm{old}}({\bf \hat{n}},z)}{\bar{T}_{b}^{\mathrm{old}}}\bar{T}_{b}^{\mathrm{new}},
     \label{eq:rescaleT}
\end{equation}
so that the new mean temperature, $\bar{T}_{b}^{\mathrm{new}}$ is given by:  
\begin{equation}
\bar{T}_{b}^{\mathrm{new}} (z)= 180h_0\frac{(1+z)^2}{E(z)}\Omega_{\mathrm{HI}}(z) mK,
\end{equation}
where $\Omega_{\mathrm{HI}}(z)$ is given by Eq. \ref{eq:omegahiska}. 

\begin{figure}
\centering
\includegraphics[trim = 0cm 0cm 0cm 0cm, width=0.45\textwidth]{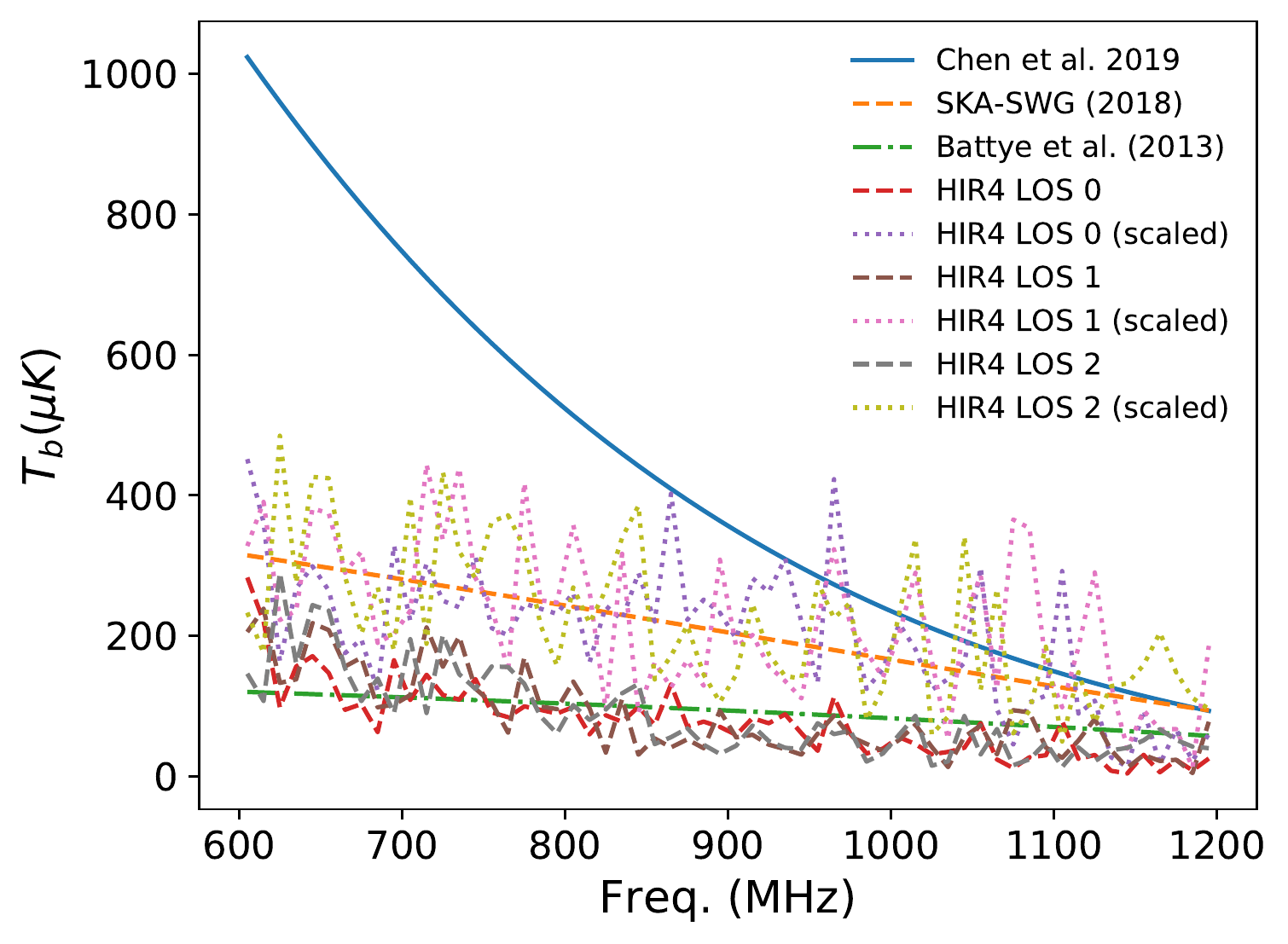}
\caption{Neutral hydrogen brightness temperature evolution with frequency. We show three different theoretical models in the literature for the mean temperature and the evolution in redshift of three different lines of sight, from the mock temperature maps for both the temperature given by the hydrogen mass from the HR4 haloes and the one rescaled following Eq. \ref{eq:rescaleT}.}
\label{fig:T_mean_freq}
\end{figure}

We can see in Fig. \ref{fig:T_mean_freq} the effect of this re-scaling in the mean temperature and in the temperature fluctuations $\Delta T_{b}= T_{b}(x)- \bar{T}_{b}$. The change in the fluctuations is needed in order to recover the right amplitude of the power spectra and the input bias from the simulation from the pure 21cm signal. We also compare our chosen approach with other definitions of brightness temperature such as \cite{2013MNRAS.434.1239B} where $\Omega_{\mathrm{HI}}$ was estimated assuming a constant mean value and only low redshift data, which tends to indicate a lower $\Omega_{\mathrm{HI}}$ or  \cite{2019JCAP...07..023C} which also consider the data compilation from \cite{2015MNRAS.452..217C} but assumes a different model based on a power law of $(1+z)$ to define the redshift evolution of $\Omega_{\mathrm{HI}} $ therefore deviating from the polynomial model that we consider in Eq. \ref{eq:omegahiska} and that is defined in \cite{2018arXiv181102743S}.

\begin{figure*}
\centering
\includegraphics[trim = 0cm 0cm 0cm 0cm, width=0.45\textwidth]{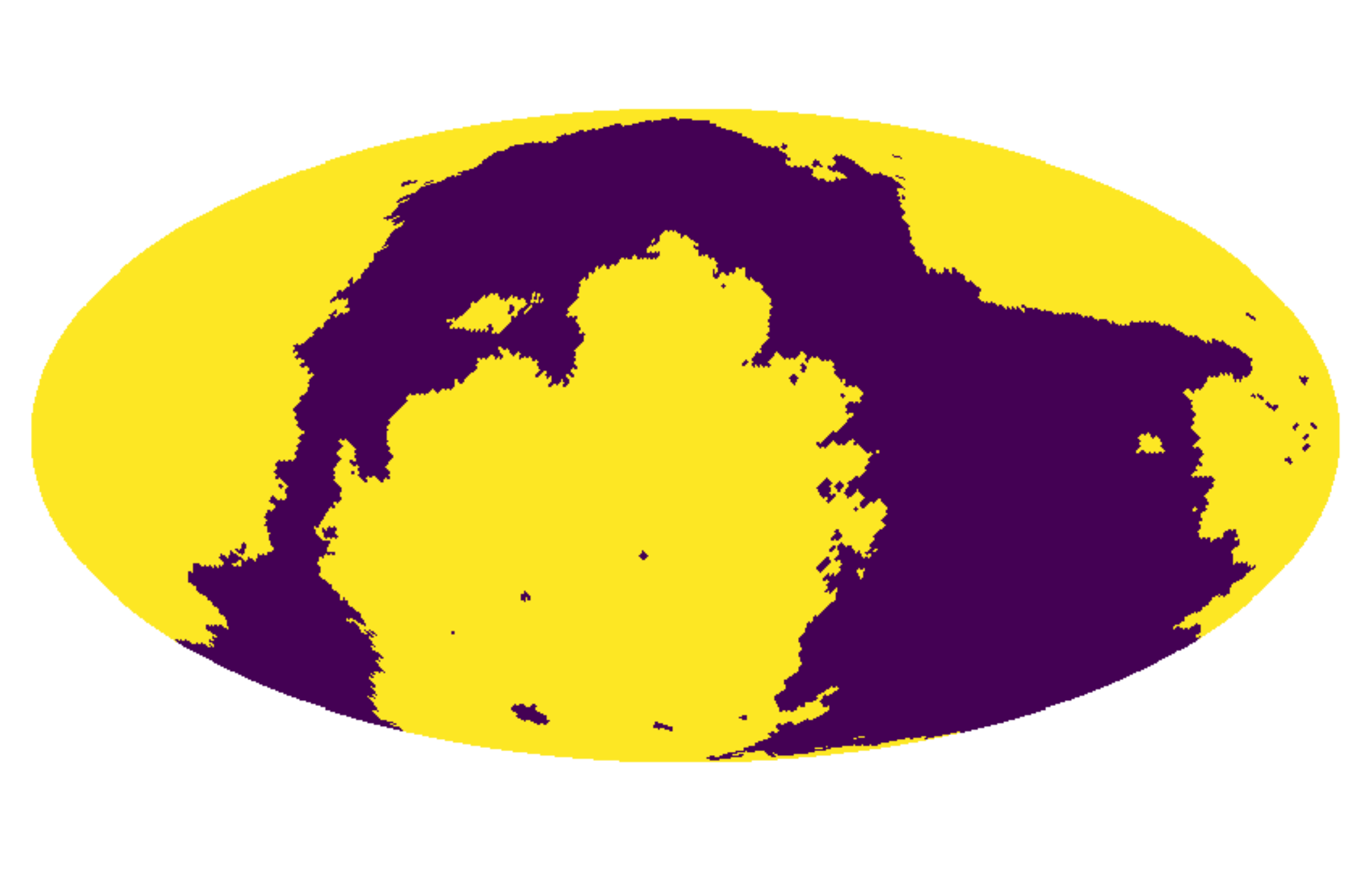}
\includegraphics[trim = 0cm 0cm 0cm 0cm, width=0.45\textwidth]{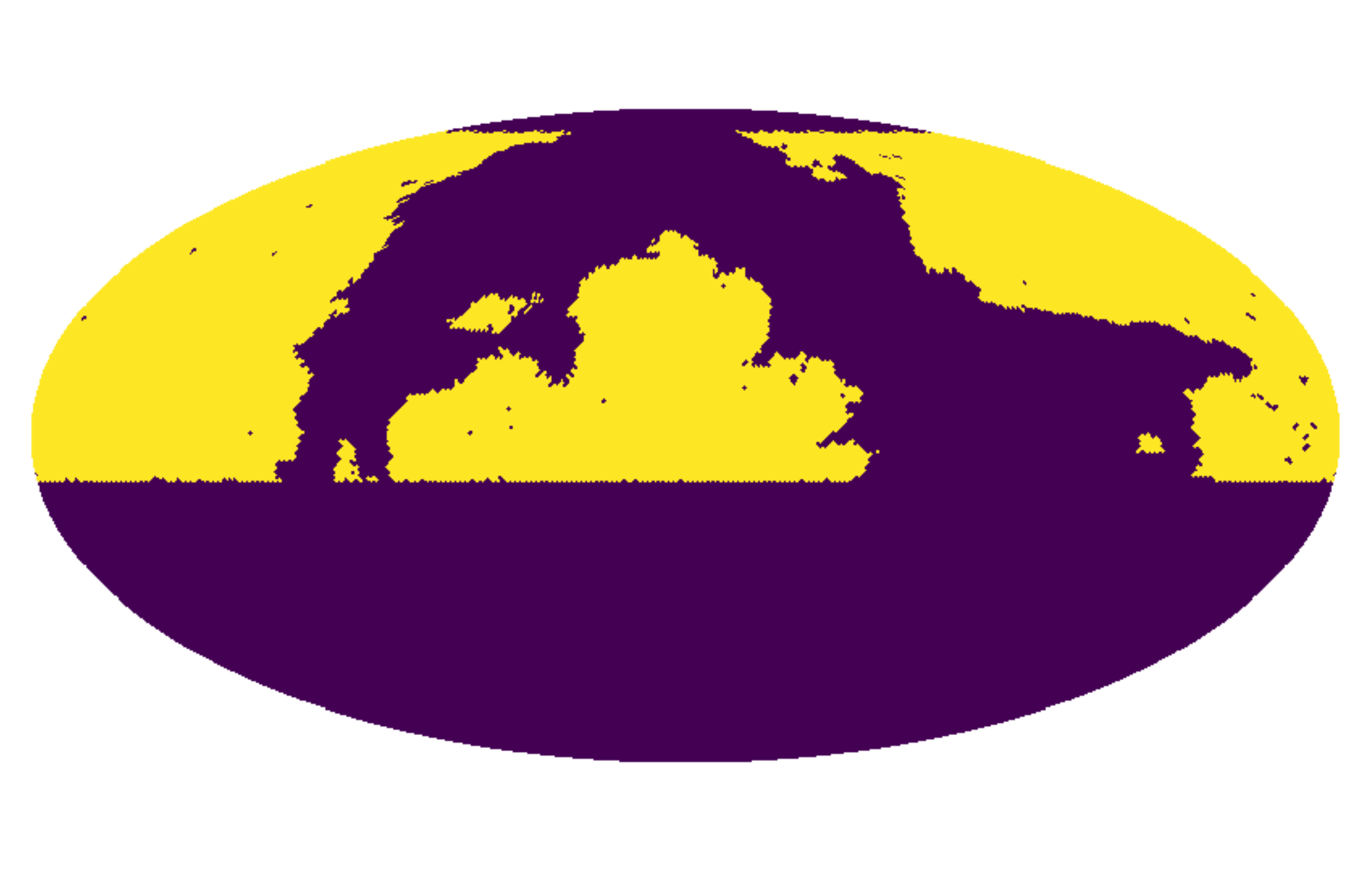}
\caption{On the left, we show the masked region in blue when removing the galactic center to reduce the foreground signal in the maps. On the right, we show the mask used when considering the Tianlai pathfinder survey noise maps as this mask includes the information of the Tianlai footprint.}
\label{fig:figmasks}
\end{figure*}

\subsubsection{Foreground maps}
\label{sec:foremaps}
Radio measurements are dominated by foreground radio emissions. In order to study the influence of foregrounds on the measured signal with our radio telescopes, we need to add or model them in our simulated catalogues. In order to do so, we have created a suite of foreground maps for each frequency bin considered in our mock catalogues. In particular, we use the updated Global Sky Model (GSM) \citep{2008MNRAS.388..247D,2017MNRAS.464.3486Z} to generate foreground maps as this method produces a good approximation to the Galactic Diffuse emission. 

GSM minimises the cost function given by a matrix decomposition and the data of 29 frequency smoothed maps with frequencies between 10MHz and 5THz using an iterative algorithm in which the initial guess is made using a PCA decomposition of 6 components of the data matrix. The GSM model maps include mostly information from five different physical mechanisms: synchrotron, free-free, CMB, warm dust and cold dust. Using the first 6 components of the PCA decomposition, we can produce a foreground temperature map, $T^{\mathrm{foreground}}_{b}(\hat{n})$, at a any given frequency within the range of the algorithm.

\subsubsection{Masking}
We can create full-sky simulations for the cosmological 21cm signal. But much of the emission from galactic foregrounds is coming from close to the galactic centre and the galactic plane. Therefore, the first step to remove the signal from foregrounds is to mask the highest intensity emission from the Milky Way. To do so, we have considered a simple procedure in which we apply a brightness temperature cut of $T_{b,mask}=8K$, and so every pixel with $T_b>T_{b,mask}$ is removed from the analysis.

We show in Fig. \ref{fig:figmasks} the masks that we use regarding the galactic emission. On the left we show in magenta the area removed from the hydrogen maps in order to estimate the observed angular power spectra, assuming that all areas of the sky are accessible, and in yellow the area that is used. On the right panel, we show the mask used for the Tianlai survey, with the same colour scheme. The yellow area is now also restricted to the footprint of the Tianlai survey, since in considering this survey, we need to apply a declination mask as Tianlai cannot access the southern ecliptic hemisphere.  Therefore, we only include the region for declinations above $\delta>-40$. We consider this mask only when using the noise maps regarding Tianlai survey that we describe in section \ref{sec:tianlai_noise}.

\subsection{Instrumental effects and Tianlai cylinder array}
\label{sec:tianlai_noise}


The last ingredient that we consider in our maps is the instrumental noise. While the 21cm cosmological signal and the foreground signal are produced by astrophysical processes, the telescopes that measure this signal have also an intrinsic thermal noise plus the extraterrestrial signal is convolved with the instrument beaming.


\subsubsection{Map making for transit radio telescopes}
Unlike a traditional radio interferometer that usually observes a small patch of sky and exploit the Fourier transform mapping between the sky and the $uv$-plane by assuming the flat-sky approximation, the Tianlai cylinder array is wide-field transit interferometers. The observable, {\it visibility}, measures different parts of the curved sky as a function of time (i.e., the azimuthal angle $\phi$). At any instant, a visibility for any particular frequency is given by
\begin{equation}
V_{ij}(\phi) = \int d^2 \hat{\bf n} B_{ij}(\hat{\bf n};\phi)  T(\hat{\bf n}) + n_{ij}(\phi)\,,
\end{equation}
where $n_{ij}(\phi)$ represents the noise term and the beam transfer function $B_{ij}(\hat{\bf n};\phi)$ is given by 
\begin{equation}
B_{ij}(\hat{\bf n};\phi) = A_i(\hat{\bf n})A^*_j(\hat{\bf n})e^{2\pi i\hat{\bf n}\cdot {\bf u}_{ij}}\,.
\end{equation}

Here, the visibility is measured by correlating the signals from a pair of feeds $i$ and $j$, located at positions $r_i$ and $r_j$ with ${\bf u}_{ij} \equiv (r_i - r_j)/\lambda$, where $\lambda$ is the observed wavelength, $\hat{\bf n}$ is the sky direction, and $A_i(\hat{\bf n})$ denotes the primary beam of feed $i$.

Recently,  a novel  ``m-mode'' formalism for the analysis of transit radio telescopes was proposed by~\cite{Shaw_2014}. It provides an easy way to linearly map wide-field interferometric data on the full sky. The measured visibilities can be written as a summation of spherical harmonic modes.  By taking into account the fact that the measured visibilities change periodically with the sidereal day (i.e., the periodicity in $\phi$), one can find a simple relation between the so-called m-mode visibilities $V_m^{ij}$ and the sky by  
\begin{equation}
V_m^{ij} = \sum_\ell B_{\ell m}^{ij}a_{\ell m} + n^{ij}_{m} \,.
\label{eq:vm}
\end{equation}
Here, $a_{\ell m}$ and $B_{\ell m}^{ij}$ denote the coefficients in the spherical harmonic expansions of the sky $T(\hat{\bf n})$ and the beam transfer function $B_{\ell m}^{ij}(\phi)$, respectively, which read
\begin{align}
T(\hat{\bf n}) &= \sum_{\ell m} a_{\ell m}Y_{\ell m}(\hat{\bf n})\,,\\
B_{ij}(\hat{\bf n};\phi) &= \sum_{\ell m} B_{\ell m}^{ij}(\phi) Y^*_{\ell m}(\hat{\bf n})\,.
\end{align}

Following Refs.~\citep{Shaw_2014,2016RAA....16..158Z}, one can then group m-mode visibilities from all baselines of the array together into a vector ${\bf v}$, and similarly group m-mode harmonic coefficients of the sky and m-mode noises for all baselines into vectors of ${\bf a}$ and ${\bf n}$, respectively. The measurement equation in Eq.~\ref{eq:vm} for each m-mode thus can be further simply rewritten in matrix form as
\begin{equation}
{\bf v= B a + n}\,,
\end{equation}
where we have expressed the beam transfer matrices in an explicit matrix notation ${\bf B}$. This equation is valid for any particular $m$ and frequency $\nu$.

Using the maximum likelihood method, the best estimate of the sky spherical harmonics coefficients from a given set of sampled visibilities for each individual $m$ and frequency $\nu$ is solved by 
\begin{equation}
\label{eq:bestmap}
\mathbf{\hat{a}} = \mathbf{ (B^{\dagger}N^{\rm -1}  B)^+ B^{\dagger} N^{\rm -1} v}\,,
\end{equation}
where the superscript $+$  represents the pseudo-inverse. Here, we assume that the instrumental noises at two different frequencies are uncorrelated and the noise follows a complex Gaussian distribution with zero mean and covariance $\left<{\bf nn^\dagger}\right> =  {\bf N}$. 

\subsubsection{Configuration of the Tianlai cylinder array}
\begin{figure}
\centering
\includegraphics[trim = 0cm 0cm 0cm 0cm, width=0.45\textwidth] {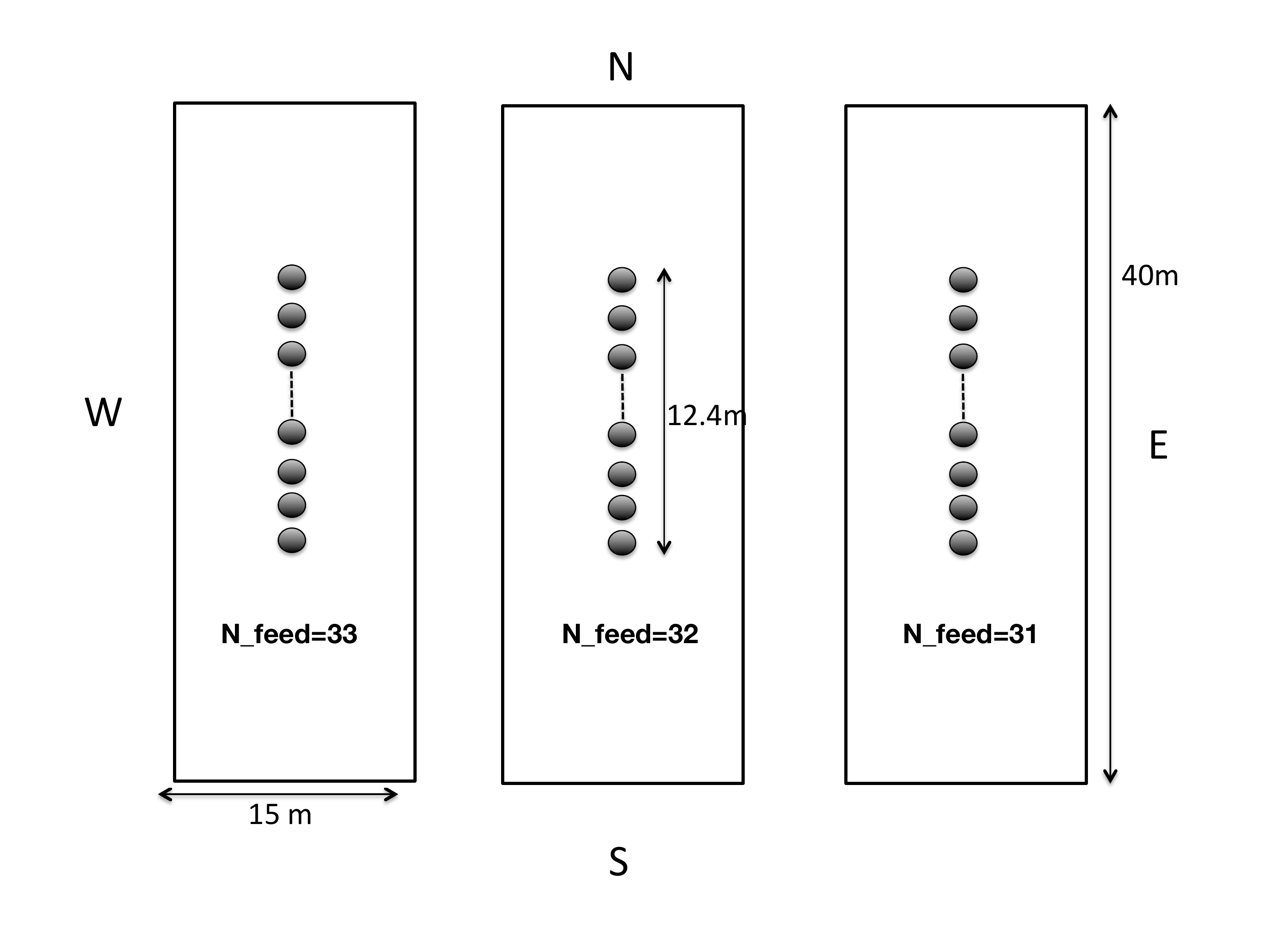}
\caption{Configuration of the Tianlai pathfinder cylinder array.  At present, the array has three adjacent cylindrical reflectors oriented in the North-South direction, each with 15 m wide and 40 m long. The three cylinders are equipped with a total of 96 dual polarization receivers, which are irregularly spaced on the cylinders. The number of feeds on each cylinder is 31, 32 and 33, respectively, spanning the same distance of 12.4 m along the North-South direction on each cylinder. The feed spacing is thus 0.388 m, 0.4 m and 0.413 m, respectively.} 
\label{fig:array} 
\end{figure}

The Tianlai Pathfinder presently consists of an array of three adjacent cylinder telescopes, located in Hongliuxia, a radio-quiet site in northwest China ($44^\circ 9'9.66''$ N $91^\circ 48'24.72''$ E). Each of the cylinders is 15 m wide and 40 m long. With wide field of view radio interferometers, the Tianlai Pathfinder is dedicated to measure the 3D maps of neutral hydrogen (so-called 21 cm intensity mapping)  of the northern sky in the Universe by surveying neutral hydrogen over large areas of the sky at low redshifts in the range of $1.03 > z > 0.78$ ($700-800$ MHz). Currently,  the three cylindrical reflectors oriented in the North-South direction, each having 33, 32, and 31 feed antennas respectively (see Fig.~\ref{fig:array}).

For the Tianlai cylinder array, by assuming uncorrelated thermal noises across all baselines and frequencies, the noise level (RMS) in units of brightness temperature is given by \citep{2001isra.book}:
\begin{equation}
\label{eq:noise_cylinder}
\sigma^N_{ij}= \left({\bf N}_m^{ij}\right)^{1/2}= \frac{T_{\rm sys}}{\sqrt{\Delta\nu \Delta t_{ij}}}\left(\frac{\lambda^2}{A_e} \right)\,,
\end{equation}
where $\Delta t_{ij}$ is the total integration time of baseline $ij$, $T_{\rm sys}$ is the system temperature, $A_e$ is the effective area of antenna, $\lambda$ is the observing frequency, and $\Delta\nu$ is the width of the frequency channel. The system temperature is the sum of the sky brightness and the analog receiver noise temperature, $T_{\rm sys} = T_{\rm sky} + T_{\rm rec}$. At the frequency of interest ($700-800$ MHz), the Tianlai array would be expected to achieve a total system temperature of $50-100$ K, and thus we assume $T_{\rm sys}$ = 50 K in this study. We also assume two full years of observation for the Tianlai pathfinder survey. The effective antenna area $A_{e}$ is calculated by $A_e\Omega = \lambda^2$, where the beam solid angle $\Omega$ is well approximated by $\Omega\simeq 0.1$ for the current Tianlai cylinder array.

By realistically simulating the noise visibilities for the Tianlai instrumental configuration, the noise sky maps, $T^{\mathrm{noise}}_{b}(\hat{n})$, for all frequencies from these visibilities are reconstructed by the above map-making process with using the maximum-likelihood solution in Eq.~\ref{eq:bestmap}.


    \label{eq:noise}
In the main analysis of this paper, we only consider Tianlai-like noise when building the HIR4 maps, but for reference we show here a comparison with a case based on SKA-MID phase 1 noise.
We can estimate the beam of the SKA survey by only considering the baseline of the telescope. The intensity mapping beam size is $\Sigma_{beam}=1.133\theta^2$. As defined in \cite{2019arXiv190401479C}, the beam resolution is giving by
\begin{equation}
    \theta_{beam}= \frac{1.22c}{\nu D_{base}}\,.
    \label{eq:theta_beam}
\end{equation}
where $D_{base}=15m$ for the SKA configuration.

We can also simulate the effect of a gaussian beam with the pure 21cm maps. In order to do so, we have smoothed the 21cm intensity mapping field in each frequency beam with a gaussian filter with a smoothing scale given by Eq. \ref{eq:theta_beam} \citep{2019arXiv190401479C}.

\subsection{Observed temperature map}
\label{sec:mapsobse}
We define the observed temperature map as the one that combines the cosmological signal from the simulation, the foregrounds and the observational noise. We define the observed temperature as:
\begin{equation}
    T^{\mathrm{obs}}_{b} (\hat{n}) = T^{\mathrm{HI}}_{b}(\hat{n}) + T^{\mathrm{foreground}}_{b}(\hat{n}) + T^{\mathrm{noise}}_{b}(\hat{n}),
\end{equation}
where $T^{\mathrm{HI}}_{b}(\hat{n})$ is the brightness temperature  of cosmological neutral hydrogen at pixel given by position $(\hat{n})$ given by Eq. \ref{eq:t_21} and with the mean temperature rescaled as given by Eq. \ref{eq:rescaleT}. Then we add the brightness temperature, $T^{\mathrm{foreground}}_{b}(\hat{n})$, from the same pixel in the foreground map, created using the GSM, at the mean frequency of the corresponding frequency bin, following prescription in section \ref{sec:foremaps}. Finally, we can add the value of the noise temperature at that angular position given by the Tianlai noise maps explained in section \ref{sec:tianlai_noise}.

%
%


\begin{figure}
\centering
\includegraphics[trim = 0cm 0cm 0cm 0cm, width=0.45\textwidth]{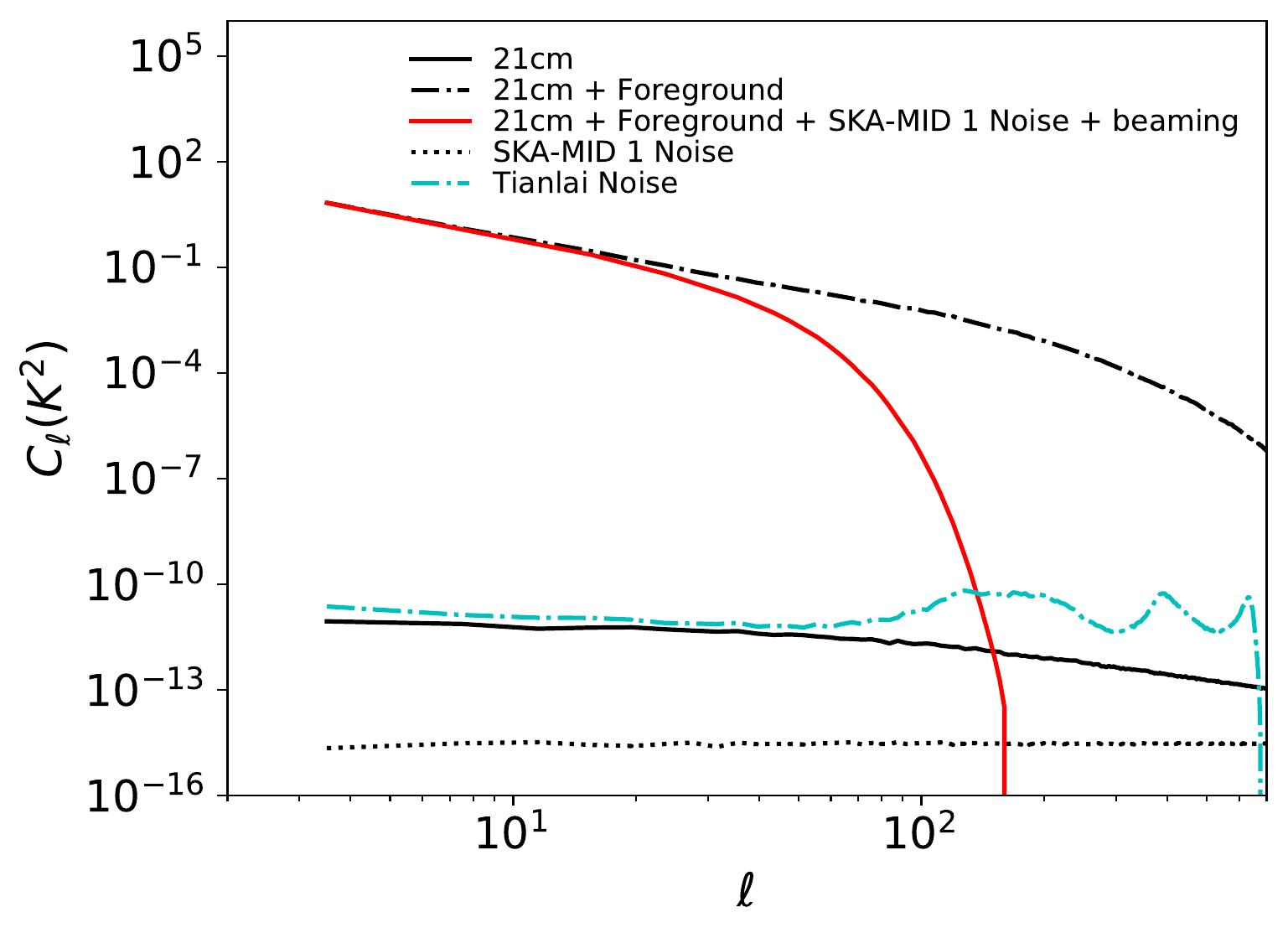}
\caption{Example of angular power spectrum for the cosmological 21cm temperature maps (black solid), maps with foreground signal (black dot-dashed), maps with foregrounds and the expected dish beam and noise maps for SKA-MID phase 1 receivers (red solid).  We also show the power spectra of the noise alone for Tianlai (cyan dot-dashed) and SKA-MID phase 1 (black dotted). The frequency band in this case is 790-800 MHz.  The SKA-MID phase 1 noise assume 10,000 hours of integration, leading to a noise variance of $\sigma = 4.4\times 10^{-5} m{\rm K}$, whereas the expected average noise for Tianlai is $\sigma =2.6\times 10^{-4} m{\rm K}$.}
\label{fig:figclsbeam}
\end{figure}

In Fig. \ref{fig:figclsbeam}, we show how the beaming affects the expected signal when we use all the information in the frequency bin 790-800 MHz. The SKA noise assume 10,000 hours of integration, leading to a noise variance of $\sigma = 4.4\times 10^{-5} m{\rm K}$, whereas the expected average noise for Tianlai given one year of integration is $\sigma =2.6\times 10^{-4} m{\rm K}$. Note that in this paper  we generate full-sky simulated noise maps for Tianlai,  assuming two full years of observations, to use in the analysis pipeline, but show that this assumption of one year on the sky gives a similar noise value to the simulated maps at large angular scales. We can see that the effect of the noise on the power spectrum  for SKA-MID phase 1  is much smaller than the effect of a Gaussian beam, given by \ref{eq:theta_beam}, over the range of multipole values that would be considered for cosmological analysis. 

\subsection{Analysis methods}

\subsubsection{Reconstruction methods}

The goal of any reconstruction method is to decompose the map into a set of signals with some different qualities or attributes. In the case of 21cm, we use reconstruction methods to split the map into the foreground part (generated locally to our Galaxy) and the cosmological part, based on the assumption that the frequency dependence of the two will be very different. In this section we describe three such methods, fast independent component analysis (fastICA), principal-component analysis (PCA), and log-polynomial fitting.

\subsection*{fastICA}
The  fastICA method assumes that the maps can decomposed into a set of signals with some non-Gaussian distribution (the foregrounds) and some Gaussian noise (the cosmological signal). This is given by
\begin{equation}
    \mathbf{x} = \mathbf{A}\mathbf{s} +\mathbf{n}\,.
\end{equation}
Here $\mathbf{x}$ is the final map, which will be split up by pixel and frequency bin,  $\mathbf{s}$ is vector of components, with an amplitude for each component that depends on the position in the map, $\mathbf{A}$ is the mixing matrix that defines how the components evolve with frequency, and $\mathbf{n}$ is the distribution of Gaussian noise. Note that the data is `pre-whitened', such that the mean in each frequency bin is removed, and only replaced during the reconstruction phase. This means that the mean temperature of each bin, which is a sum of the mean of the 21cm cosmological signal and the mean of the foreground emission at that frequency, cannot be reconstructed by this process, and only the distribution of fluctuations can be separated.

In our implementation, the non-Gaussian components should correspond to the foregrounds, which should be well behaved and continuous in frequency space. In contrast,  the intensity of the cosmological 21cm emission depends on the mass of neutral hydrogen present in the `voxel', which is a stochastic quantity with a Gaussian distribution, and so resembles the noise in a fastICA reconstruction process. As such, the reconstructed map will be given by
\begin{equation}
    \label{eqn:reconstructing}
    \mathbf{x}_{\rm recon} = \mathbf{x}_{\rm data} - \mathbf{A}\mathbf{s}\,,
\end{equation}
where $\mathbf{x}_{\rm recon}$ is the reconstructed map, ordered by pixel and frequency bin, which includes both the cosmological 21cm distribution and the Gaussian noise from the instrument.

The method used to estimate $\mathbf{s}$ the vector of components, and $\mathbf{A}$ the mixing matrix, is very similar to that described in \citet{2012MNRAS.423.2518C,FastICA} and other, previous works.  Using the implementation of fastICA as part of the {\tt scikit-learn} python machine learning package \citep{scikit-learn} we maximise the negentropy, defined by $J(y)=H(y_{\mathrm{gauss}})-H(y)$, assuming the negentropy is approximated by a $\log\cosh(y)$ function. The negenetropy functions as a measure of distance from gaussianity, and so maximising it with respect to the components should remove the foreground signal, leaving behind the Gaussian cosmological signal.

\subsection*{PCA}
For the principal-component analysis (PCA) method, we again work with ``pre-whitened'' data, where the mean of the simulated data has been subtracted. From this we  compute the data covariance matrix, looking at the covariance between different bins in frequency space. We then compute the eigenvectors and eigenvalues of this frequency-frequency covariance matrix. Since the foregrounds dominate the power in  maps at all frequencies, they will dominate the eigenmodes with the highest eigenvalues. Also, the foregrounds are expected to have smooth frequency structure, so that they could be described by just a few smooth frequency eigenvectors. With these reasons, finally we project out the principal components with the largest eigenvalues in frequency space from every spatial pixels to obtain foreground cleaned maps. However, correlations in frequency space can also be slightly generated cosmologically, and so again this foreground removal approach may also have the effect of removing the 21 cm signal from the power spectrum over all scales.

Specifically, following~\cite{2008MNRAS.388..247D}, we reshape the three-dimensional observed data into an $N_\nu\times N_\theta$ matrix ${\bf x}$, where $N_\theta$ contains all two-dimensional spatial pixels (the same as in the fastICA approach). The empirical $\nu-\nu$ covariance of the data is
\begin{equation}
\mathbf{C} = \frac{\mathbf{xx}^T}{N_\theta}\,.
\end{equation} 
By using the PCA analysis, the matrix ${\bf C}$ can be decomposed into  ${\bf C = U\Lambda U}^T$, where ${\bf \Lambda}$ is diagonal and contains the eigenvalues in descending order and ${\bf U}$ is an orthogonal matrix whose columns are the eigenvectors (i.e., the principal components). Now we define the deprojection matrix, ${\bf \Pi} = {\bf I- USU}^T$. Here ${\bf I}$ is the identity matrix and ${\bf S}$ is a selection matrix with 1 along the diagonal for modes to be removed and 0 elsewhere.  We can then apply ${\bf \Pi}$ to our map along each line of sight 
\begin{equation}
\mathbf{x}_{\rm recon} = \mathbf{x}_{\rm data} \left[{\bf I- USU}^T \right]\,,
\end{equation}
in order to project out the selected principal components which are significantly dominated by foregrounds.

\subsection*{log-polynomial fitting}
For the log-polynomial fitting, we do not use the pre-whitened field
but take the raw combined map ${\bf x}$ and try the linear least-square fitting with an $n$-th order polynomial,
\begin{equation}
\log T(\hat{n},\,\nu)=\sum_{j=0}^{n}s_{j}(\hat{n})\left(\log\nu\right)^{j}\,,\label{eq:logpoly}
\end{equation}
at every direction $\hat{n}$. We do not consider the noise covariance
matrix at this point, such that the fitting becomes equivalent to
\begin{equation}
{\bf y}={\bf A}{\bf s}\,,\label{eq:polymatrix}
\end{equation}
at every direction $\hat{n}$, with ${\bf y}\equiv\{\log T(\hat{n},\,\nu_{i})\}$,
$A_{ij}\equiv\left(\log\nu_{i}\right)^{j}$ and ${\bf s}\equiv\{s_{j}(\hat{n})\}$.
The best-fit parameter set is then given by the estimator
\begin{equation}
\hat{\bf{s}}=({\bf A}^{T}{\bf A})^{-1}{\bf A}^{T}{\bf y}\,,\label{eq:polyestimator}
\end{equation}
where the superscript $T$ denotes the transpose. The reconstructed
map is then given by ${\bf x}_{{\rm recon}}={\bf x}_{{\rm data}}-{\bf A}{\bf \hat{s}}$
just as in Eq. \ref{eqn:reconstructing}. This is obviously not the best practice that considers the property of the noise, but is one that just relies on the smoothness of the foreground. One can of course follow the procedure by \citet{2008MNRAS.388..247D} with the noise covariance matrix ${\bf N}\equiv\left\langle {\bf n}{\bf n}^{T}\right\rangle $
for a better estimator,
\begin{equation}
\hat{s}=({\bf A}^{T}{\bf N}^{-1}{\bf A})^{-1}{\bf A}^{T}{\bf N}^{-1}{\bf y},\,,\label{eq:polyestimator_better}
\end{equation}
once ${\bf N}$ properly reflects the instrumental noise and the cosmic
signal. Because the log-polynomial fitting is already found to be
out-performed by PCA in a wide range of frequency \citep{2008MNRAS.388..247D}, we simply use Eq. (\ref{eq:logpoly}-\ref{eq:polyestimator})
in this work to illustrate its relative power.

\subsubsection{Angular power spectrum}

In order to study the cosmological information encoded in our maps, we decompose the distribution of intensity in a certain basis set (in this case spherical Bessel functions). If the continuous intensity field in a particular direction $T(\vec{\theta})$ is Gaussian and randomly distributed, then it can be decomposed into its multiple moment using spherical harmonics $Y_{\ell m}$
\begin{equation}
a_{\ell m} = \int d\vec{\theta} Y^*_{\ell m} T(\vec{\theta})\,.
\end{equation}
Assuming an isotropic universe, we get the power spectrum from the autocorrelation function,
\begin{equation}
\langle a^*_{\ell m}a_{\ell' m'}\rangle = \delta_{\ell \ell'}\delta_{mm'}C_{\ell}\,.
\end{equation}
Since the spherical Bessel functions are dimensionless, the spherical harmonic coefficients $a_{\ell m}$ must have units of intensity or temperature per unit area, and the power spectrum $C_\ell \propto T^2$.

To measure the angular power spectrum we used the {\tt NaMaster}\footnote{Downloaded from \url{https://github.com/LSSTDESC/NaMaster}.} code \citep{NaMaster},  which uses the pseudo-Cl (aka MASTER) approach, including the effect of the sky mask.

\subsubsection{Covariance matrix}
\label{sec:cov}
Finally, we need to include the measurement errors on the angular power spectra in order to constrain the cosmological parameters. As we are focusing on the linear scales in this paper, we assume that the density field is linear and described by a Gaussian distribution, in order to define the covariance matrix. When considering a full sky map, it might be possible to assume that the different modes are not correlated, giving the standard relation that 
\begin{equation}
    \mathrm{Cov}(C_\ell,C_{\ell\prime})= \mathrm{Var}(C_\ell)\delta_{\ell,\ell\prime}\,,  
\end{equation}
where the cosmic variance component of the angular power spectra variance is given by
\begin{equation}
    \mathrm{Var}(C_\ell) =  \frac{2C_\ell^2}{N(\ell)}\,
    \label{eq:gauss_cov}
\end{equation}
where $N(\ell)=(2\ell+1)\Delta \ell f_{\mathrm{sky}}$ is the number of modes for each multipole $\ell$ \citep{2007MNRAS.381.1347C,2011MNRAS.414..329C,2012MNRAS.427.1891A}. The variance depends therefore on the fraction of sky used for the cosmological analysis, $f_{\mathrm{sky}}=A_{\text{survey}}/A_{\text{sphere}}$, and on the amplitude of the multipole bins $\Delta \ell$ used to measure the angular power spectra. As we include the effect of the angular mask by drawing theoretical realizations of the density field, we assume $f_{sky}=1$ when using Eq. \ref{eq:gauss_cov} to get the first estimates of the cosmological parameters before re-doing the analysis with a more realistic covariance matrix.

\begin{figure}
\centering
\includegraphics[trim = 0cm 0cm 0cm 0cm, width=0.45\textwidth]{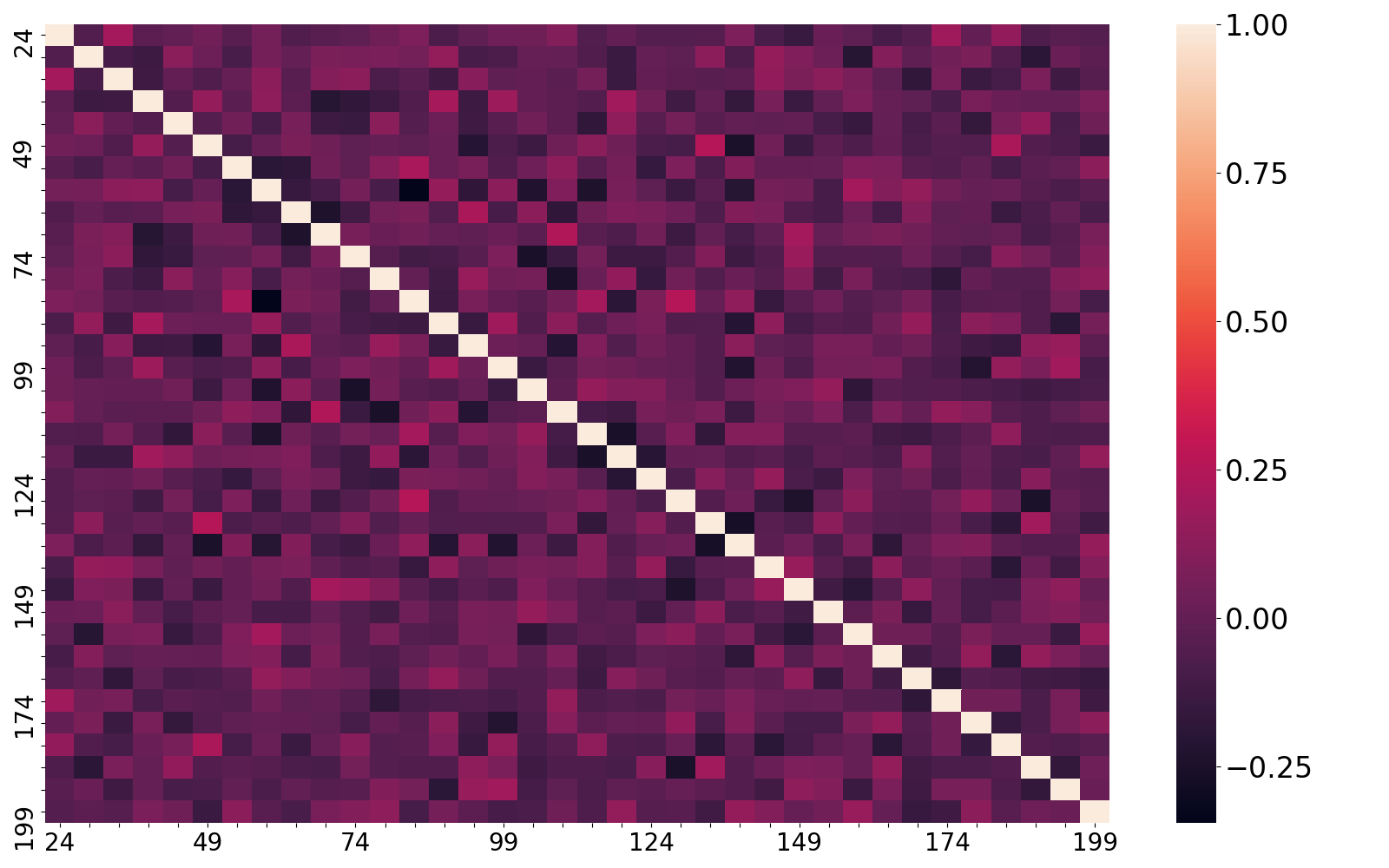}
\caption{Correlation matrix $r=C_{ij}/(\sqrt{C_{ii}C_{jj}}$for the different $\ell$-values considered in our cosmological analysis for the $f=790-800$ MHz frequency bin.}
\label{fig:HImapnside128}
\end{figure}

However, we will be removing part of the sky to reduce the effect of foregrounds, and so mode-mode coupling in the cut sky can lead to non-zero off-diagonal elements in the covariance matrix (e.g. \citet{2005MNRAS.360.1262B}). To estimate the full covariance matrix for the intensity map, including cut-sky effects, we would need a large number of simulated skies, each a different realisation of an n-body simulation. Horizon Run 4 is only a single realisation, and we don't have multiple simulations of equal volume and resolution available. To get around this problem, we create a large number of Gaussian simulated skies, $n_{\mathrm{realizations}}=100$, with the same measured power spectrum as the HR4 intensity map. We then used NaMaster to measure the power spectrum and cross-correlations between $\ell$-values across the ensemble of Gaussian realisations to estimate the co-variance. We show an example of the correlation matrix in Fig. \ref{fig:HImapnside128} of the $f=790-800$ MHz frequency bin angular power spectrum.

\subsubsection{Cosmological parameter estimation}

Once we have the measured angular power spectrum and estimated the covariance matrix, we can use this to estimate values of the cosmological parameters. The results from this parameter fitting can be used to indicate if there is any bias/offset that has been introduced by the reconstruction methods. In this case we choose the amplitude scaling parameters, the linear bias $b$, the growth rate $f$ and the background 21cm temperature $\bar{T}_{21}$. We fix the other cosmological parameters to the values specified in section \ref{HR4sims}. 

To speed up the analysis, we decompose the angular power spectrum into the term that depends only on the bias, the term that depends only on the RSD effect, and the cross-term between the two. Following the approach of \cite{2012MNRAS.427.1891A,2014MNRAS.445.2825A}, we can reconstruct any angular power spectrum in the parameter space of $\{b,f,\bar{T}_{b}\}$ through the following equation
\begin{equation}
\label{eqn:cl_recon}
    C_{\ell} = \bar{T}_{21}^2\left[ b^2C_{\ell}^{\rm no~rsd} +f^2C_{\ell}^{\rm no~bias} + 2bfC_{\ell}^{\rm cross}\right]\,.
\end{equation}
This approach assumes no growth or bias evolution through the redshift bin, which is justified given the very thin redshift slicing that can be performed. It also fixes the overall amplitude of the cosmological density field $\sigma_8$.

We see from Eq. \ref{eqn:cl_recon} that the background 21cm temperature parameter is completely degenerate with a combination of the growth and bias parameters. Of these three parameters then, only two can be independently measured by the 21cm auto-correlation angular power spectrum. (This degeneracy can be broken with cross-correlations with other tracers, but we leave that discussion for the future.) We fix $\bar{T}_{b}$, and  measure the combinations $\bar{T}_{b}b$ and $\bar{T}_{b}f$. 

We use Bayes' theorem to estimate the posterior distribution of the free parameters $\theta$ of our model, given the mock data generated from HR4 $D$. Bayes' theorem is given by
\begin{equation}
    P(\theta|D) = \frac{P(D|\theta)P(\theta)}{P(D)}\,.
\end{equation}
Here $P(\theta|D)$ is the posterior, $P(\theta)$ is the prior, and $P(D|\theta)$ is the likelihood. $P(D)$ is the evidence, which here is an unimportant overall normalisation factor.

For the likelihood we use the $\chi^2$ with Gaussian errors, such that
\begin{equation}
    \chi^2 = R^T C^{-1} R\,,
\end{equation}
where $R$ is the array of the residual, the difference between the theoretical value from Eq.~\ref{eqn:cl_recon} and the data. We weight this difference with the inverse of the covariance matrix $C=\mathrm{Cov}(C_\ell,C_{\ell\prime})$, defined in section \ref{sec:cov} and it is constant when sampling the space of parameters. Finally, the prior is set such that both $b$ and $f>0$, with some large upper limit.

We use the affine-invariant ensemble sampler, known as MCMC hammer and described in \citet{Foreman_Mackey_2013},\footnote{The code emcee can be found at \url{https://github.com/dfm/emcee}.} 
to sample the parameter space.

\section{Results}
\label{sec:results}

\subsection{Maps}
We have created full sky maps in the $700-800$ MHz frequency range. We have created pure 21cm maps for three different bandwidths, $df=2.5, 5$ and $10$ MHz. This matches the expected frequency range of the Tianlai survey. Although we can go up to frequency bands of $df=1$ MHz, we have decided to test this three different configurations to test our simulated maps, our pipeline and the growth rate of structure test with different layers of systematic errors.

\begin{figure*}
\centering
\includegraphics[trim = 0cm 0cm 0cm 0cm, width=0.9\textwidth]{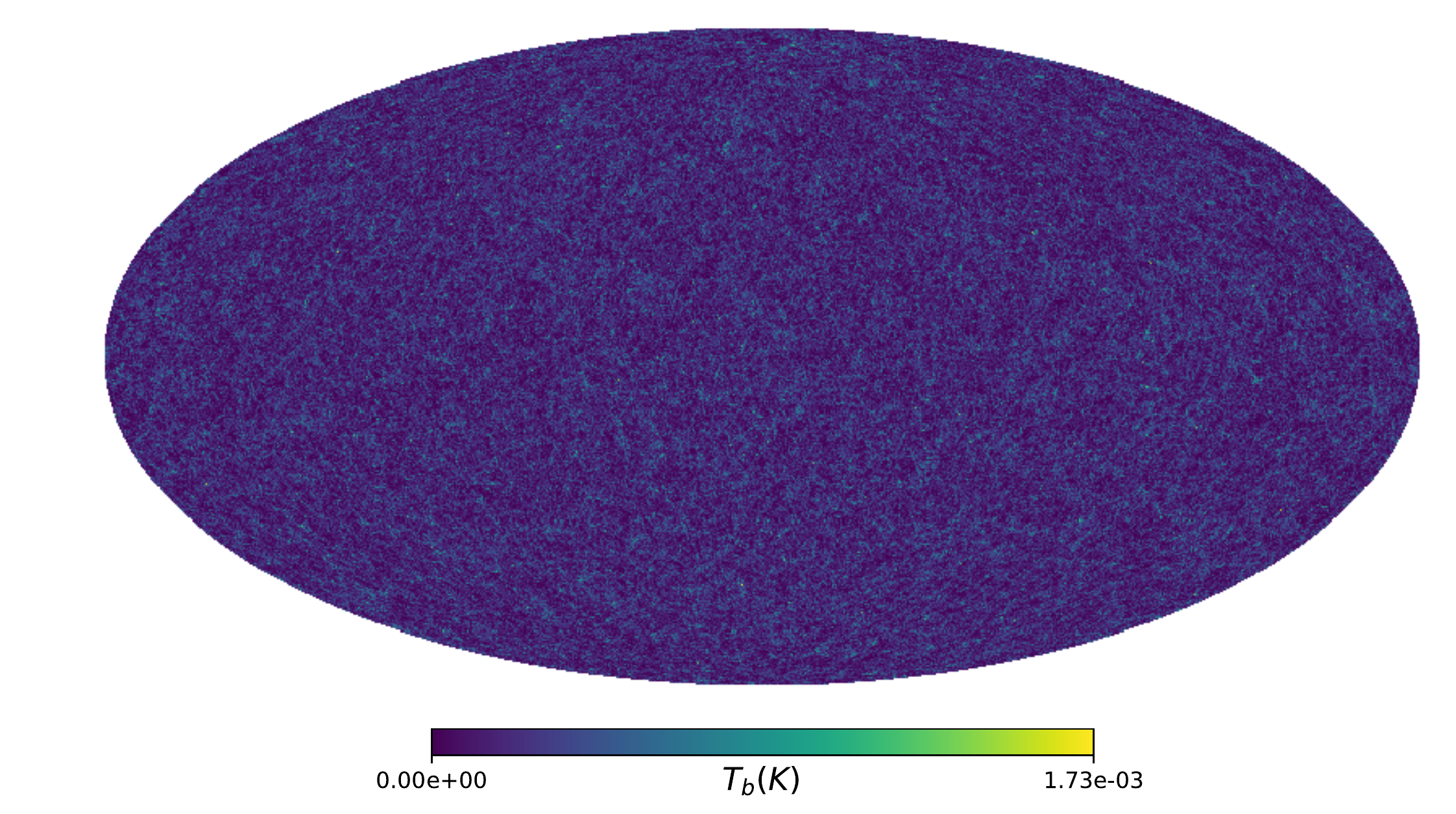}
\caption{Neutral hydrogen map for frequency $f=790-800$ MHz and $n_{side}=512$ generated from the full sky HR4 halo catalogue.}
\label{fig:HImapnside512}
\end{figure*}

\begin{figure*}
\centering
\includegraphics[trim = 0cm 0cm 0cm 0cm, width=0.45\textwidth]{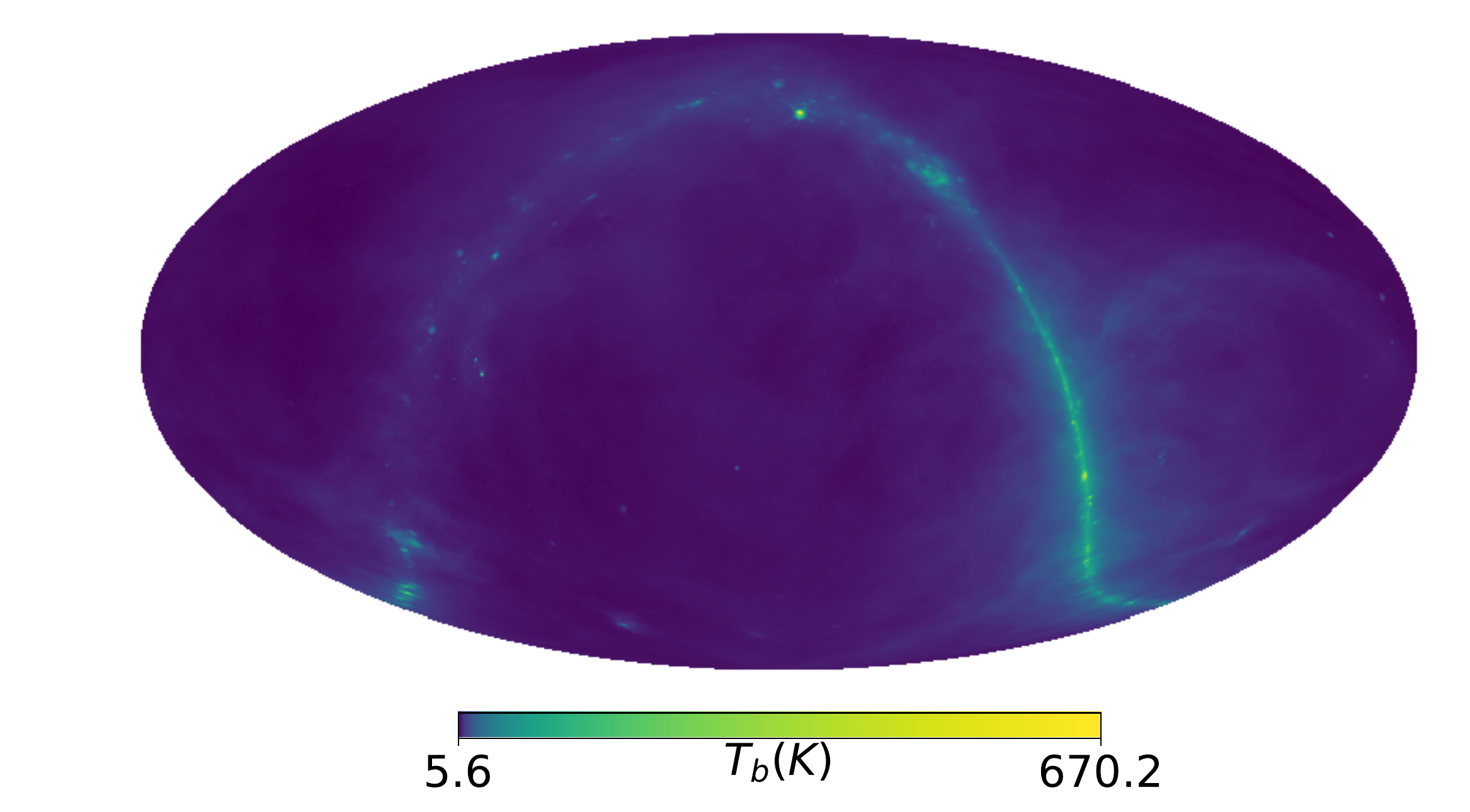}
\includegraphics[trim = 0cm 0cm 0cm 0cm, width=0.45\textwidth]{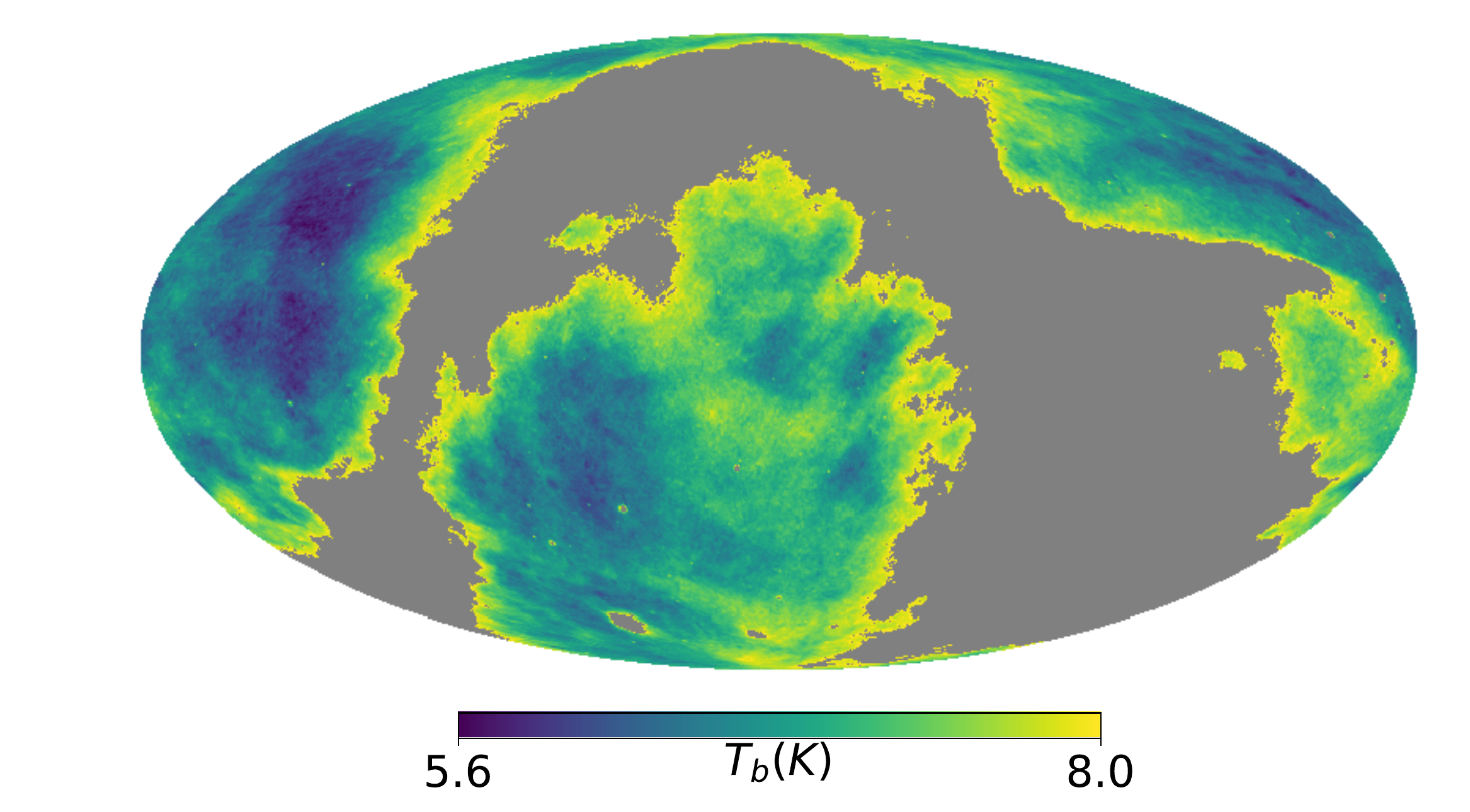}
\caption{On the left, neutral hydrogen map for frequency $f=790-800$ MHz and $n_{side}=512$ generated from the full sky HR4 halo catalogue and combined with a foreground map generated using the GSM model at $f=795$ MHz. On the right, same map is masked with the MW mask shown in the left panel of Fig. \ref{fig:figmasks}.}
\label{fig:HImapnsidefore512}
\end{figure*}

\begin{figure*}
    \centering
    \begin{tabular}{ccc}
    \includegraphics[width=0.3\textwidth]{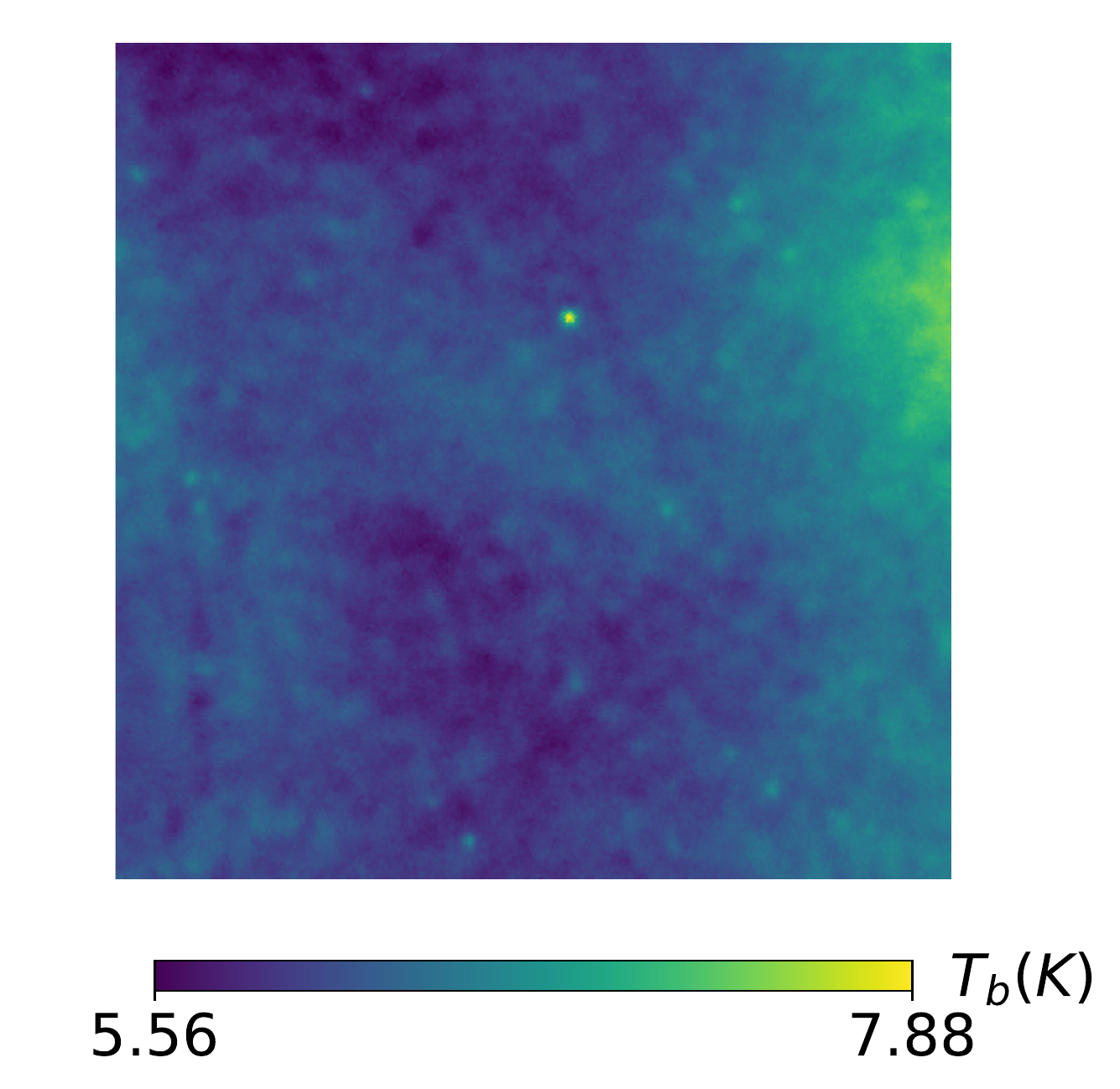}&\includegraphics[width=0.305\textwidth]{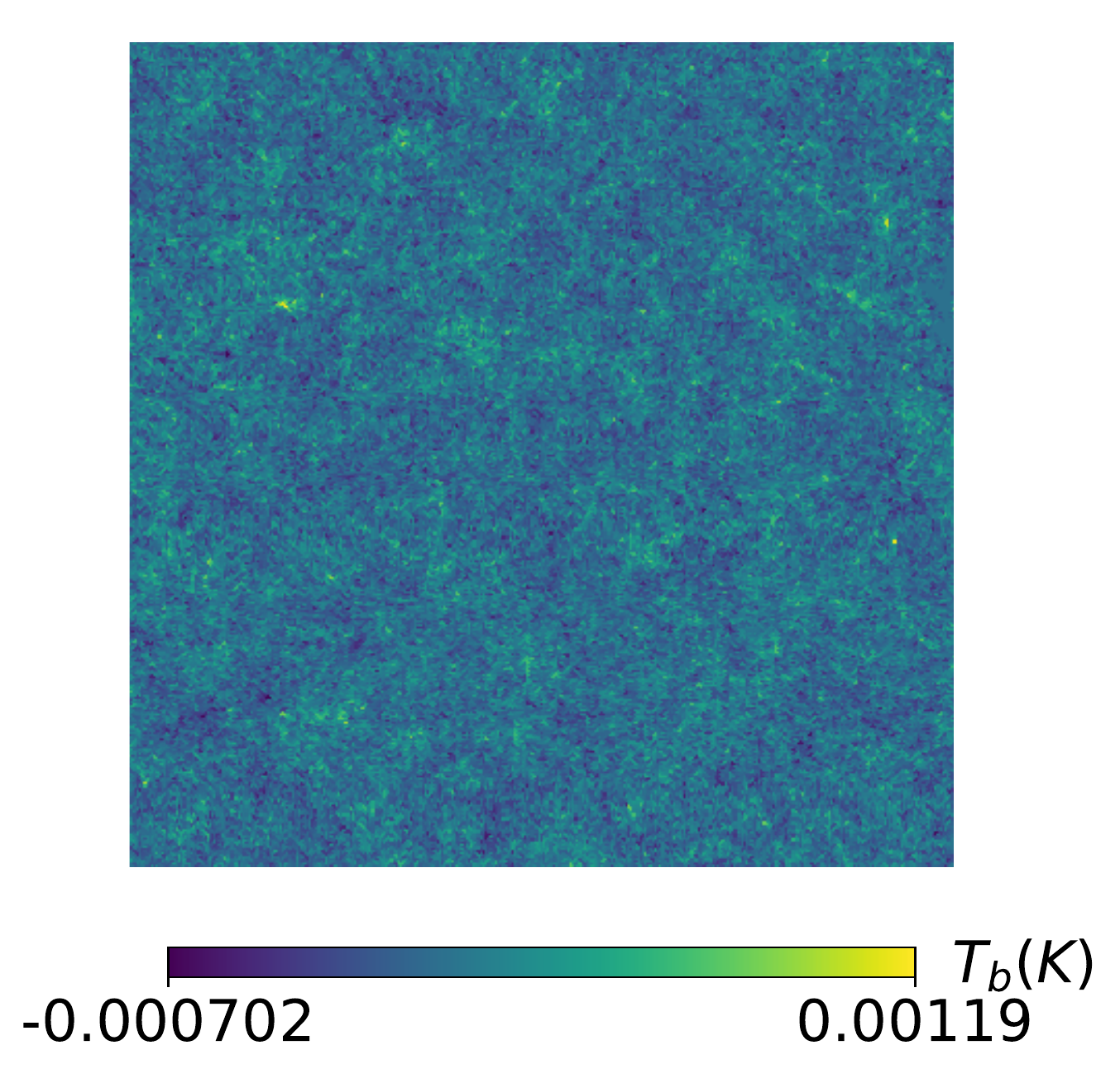}&\includegraphics[width=0.32\textwidth]{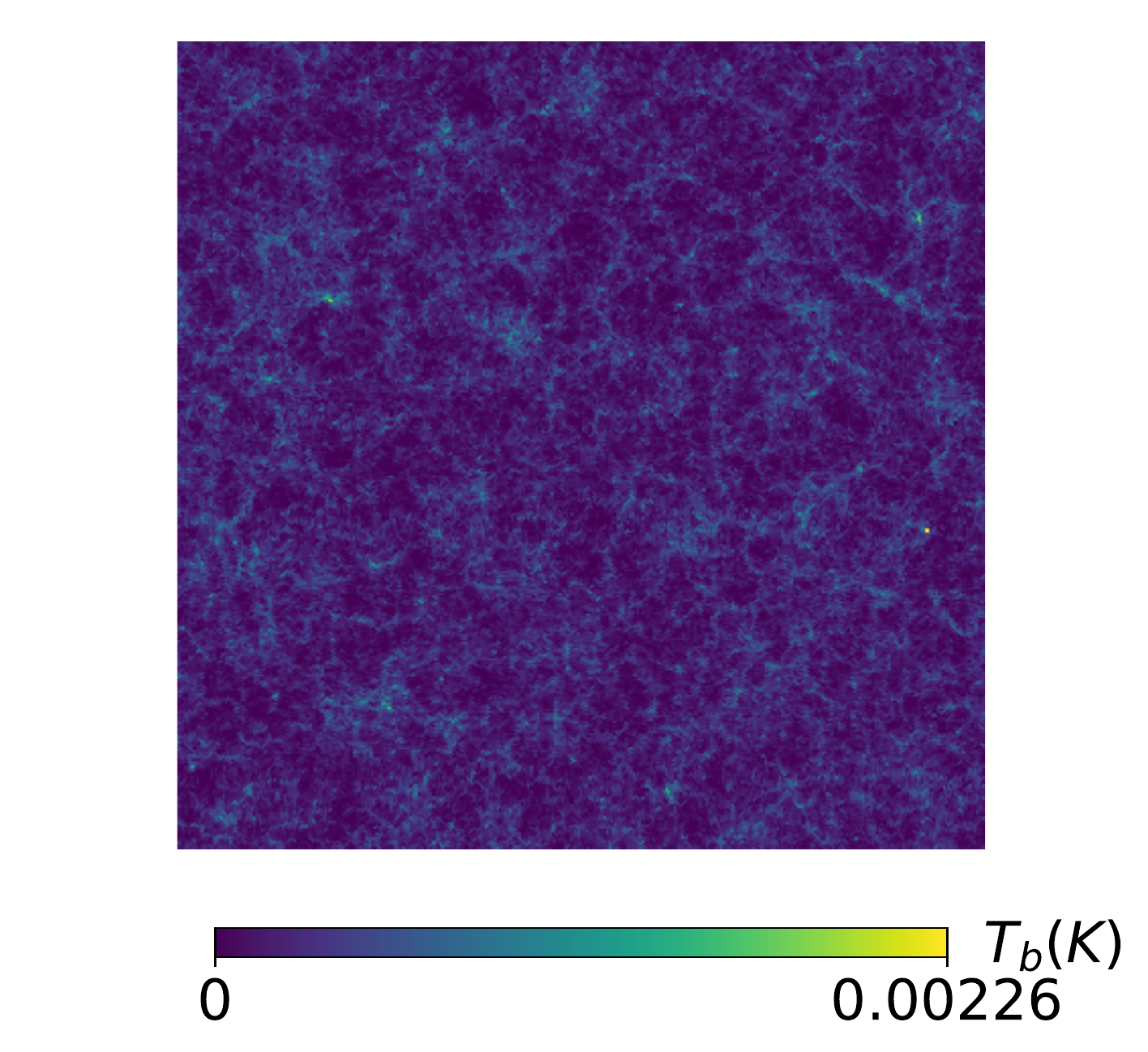}
    \end{tabular}
    \caption{On the left, we show the observed map that includes the cosmological signal and the radio foreground signal. On the center, the fastICA recovered map is shown after removing first 2 non-gaussian components. On the right panel, we show the original cosmological map. We can see how the central and the right panels are similar.}
    \label{fig:maps_diff}
\end{figure*}

In Fig. \ref{fig:HImapnside512} we show a 21 cm map for the frequency bin of $f=790-800$ MHz as an example of our mock pure 21cm maps. This map contains the cosmological information encoded in the neutral hydrogen as tracer of matter in the redshift range $0.775<z<0.797$ given by the frequency band in which we have chosen to select galaxies when splitting the total frequency range in 10 bins with bandwidths of $df=10$ MHz. We must notice that we are capable of producing full sky maps because the HR4 halo catalogue is a full sky simulation up to redshift $z=1.5$.

However, we show the contrast with a map that includes the foreground emission in Fig. \ref{fig:HImapnsidefore512}. We see that the amplitude of the foreground signal is significantly higher than the cosmological signal. The first attempt of foreground removal we applied consisted on applying the Milky Way cut to remove most of the Galactic Diffuse emission. This is shown in the right panel of Fig. \ref{fig:HImapnsidefore512}.

In Fig. \ref{fig:maps_diff}, we show a comparison between the different components that we consider on our maps. On the right panel, we show a patch of the map shown in Fig. \ref{fig:HImapnside512} which corresponds to the brightness temperature of neutral hydrogen in the frequency bin $790-800$ MHz. On the left panel, we show what corresponds to an observed map, without considering receiver noise. This map only includes the information from the cosmological neutral hydrogen brightness temperature and the foreground signal in the same redshift range given by the GSM. As can be seen, no structure can be distinguished when the foregrounds are added. In the central panel, we show the reconstructed brightness temperature map when fastICA has been applied to the map on the left panel and considering two  components. We can recognise the structure on the right panel in the middle panel. 

\begin{figure}
    \centering
    \includegraphics[width=0.45\textwidth]{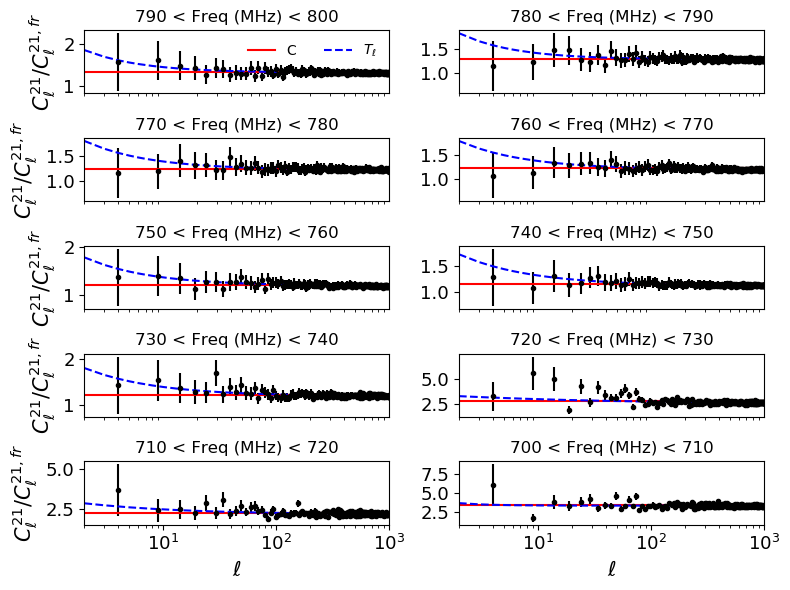}
    \caption{We show the transfer function for the $10$ bins in the $df=10$ MHz configuration, from the lower redshift bin on the top left to the higher redshift bin in the bottom right. The black circles correspond to the ratio between $C_{\ell}^{21}/C_{\ell}^{21,fr}$ while the  dashed line corresponds to the function given by Eq.  \ref{eq:fit_transfer} and the best fit values of $\ell_\star$ and $C$ for each frequency bin. The solid line corresponds to the best fit value of $C$ in each redshift bin.}
    \label{fig:transfer_function}
\end{figure}

\begin{figure*}
    \centering
    \includegraphics[width=0.9\textwidth]{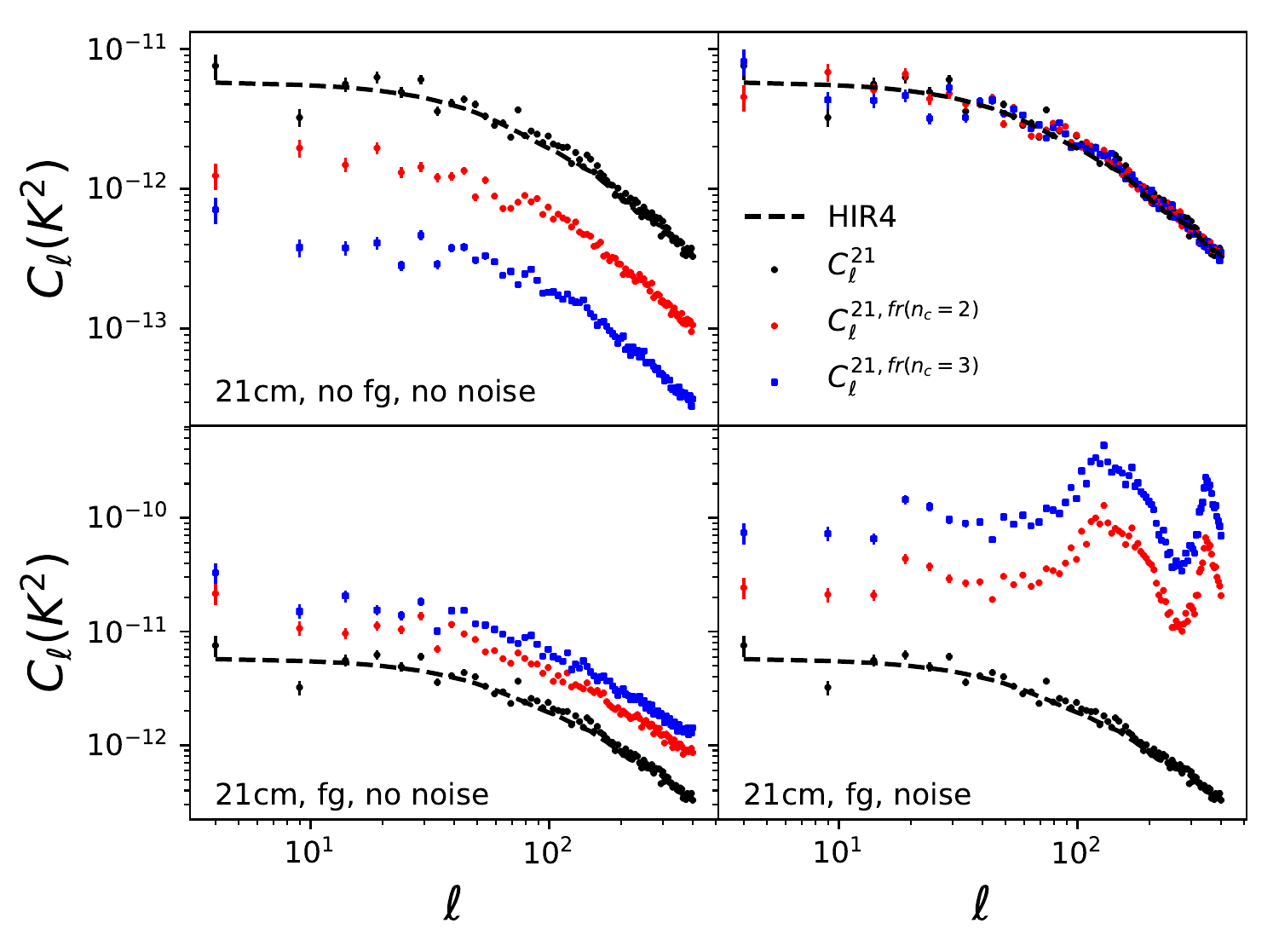}
    \caption{We show the angular power spectra for the $f=700-710$ MHz frequency bin. On the top left, we show a comparison between the pure 21cm signal and the angular power spectra corresponding to fastICA foreground removal maps for two and three components. On the top right, we apply the transfer functions to recover the original cosmological information. On the bottom left, we apply the transfer function to the maps with foregrounds, after foreground removal. Finally in the bottom right we include the Tianlai pathfinder noise map. Black dashed line corresponds to the simulation input.} 
    \label{fig:df10_cls_rec}
\end{figure*}

\begin{figure*}
    \centering
    \includegraphics[width=0.9\textwidth]{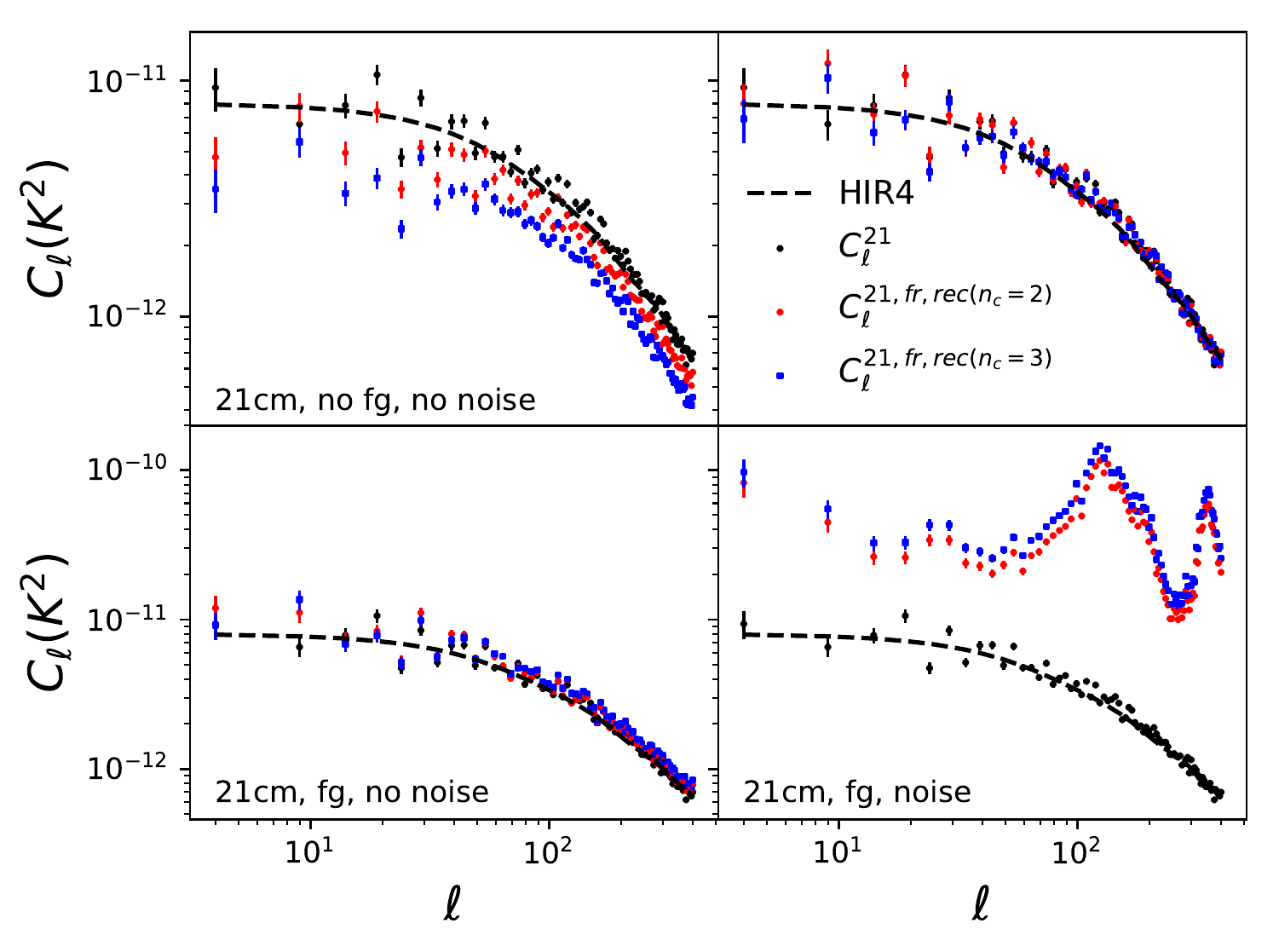}
    \caption{In this figure, we show the angular power spectra for the $f=700-705$ MHz frequency bin. The top left panel shows a comparison between the pure 21cm signal and the angular power spectra corresponding to fastICA foreground removal maps for two and three components. On the top right, we show the reconstruction with the transfer functions to recover the original cosmological information. On the bottom left, we show the reconstruction with the transfer function for the maps with foregrounds. Bottom right panel includes the Tianlai pathfinder noise map and the simulation input (dashed line).} 
    \label{fig:df5_cls_rec}
\end{figure*}

\begin{figure*}
    \centering
    \includegraphics[width=0.9\textwidth]{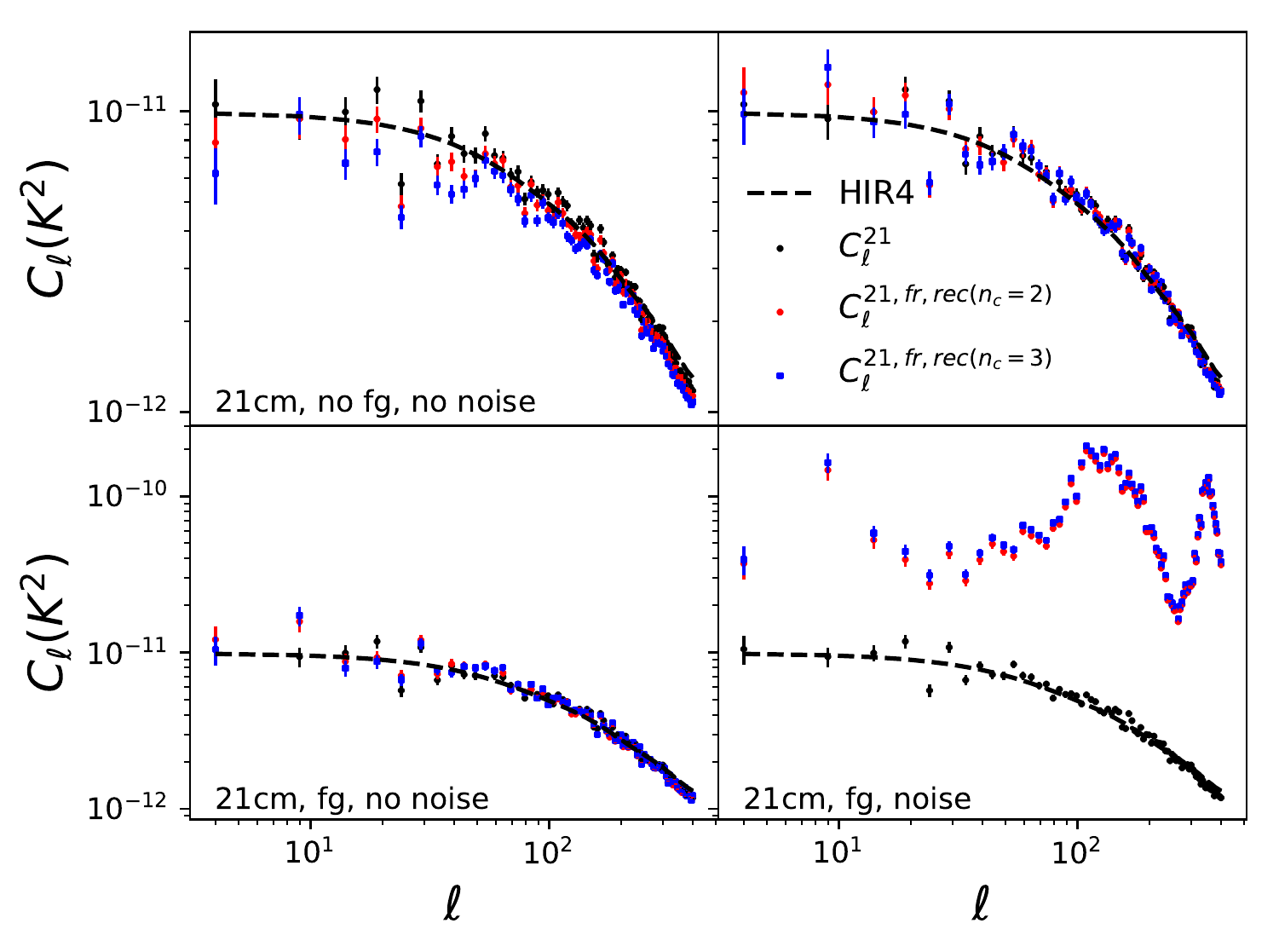}
    \caption{Same as Fig. \ref{fig:df10_cls_rec} and \ref{fig:df5_cls_rec} for the $f=700-702.5$ MHz frequency bin. On the top left panel, we show a comparison between the pure 21cm signal and the angular power spectra corresponding to fastICA foreground removal maps for two and three components with the correction function applied in the top right panel. We show the same on the bottom left but when the maps include the radio foregrounds. Finally in the bottom right we include the Tianlai pathfinder noise and all the plots include the input value from the simulation (dashed line).} 
    \label{fig:df2p5_cls_rec}
\end{figure*}

\subsection{Reconstructed angular power spectra}
We define the reconstructed angular power spectra as the ones given by the maps obtained after removing foregrounds, first with a milky way mask, and secondly applying a foreground removal algorithm.

\subsubsection{Transfer functions}
\label{sec:transfer_function}
The main goal of foreground removal is to reconstruct the original cosmological information.
When we compare the reconstructed angular power spectra $C_{\ell}$ after foreground removal with the angular power spectra from the cosmological maps, we find that the foreground removal process removes partially cosmological information as seen in Fig. \ref{fig:maps_diff}.
 
The approach we have used in order to try to reconstruct the removed cosmological information is to define a transfer function, $T_{\ell}$:
\begin{equation}
C_{\ell}^{21} = T_{\ell}C_{\ell}^{21,fr}
\label{eq:transfer_function}
\end{equation}
that is calibrated using the pure cosmological 21cm map, $C_{\ell}^{21}$ and the same map once we run our foreground removal ({\it fr}) method on it, $C_\ell^{21,fr}$. The procedure is simple, we run the foreground removal technique on the simulated cosmological 21cm maps in order to measure $C_{\ell}^{21,fr}$. Then, we fit the ratio $C_{\ell}^{21}/C_{\ell}^{21,fr}$    
to the following functional form of a transfer function:
\begin{equation}
    T_{\ell} = \exp^{-\ell_\star\log{\ell}} + C
    \label{eq:fit_transfer}
\end{equation}
given by a power law in the large scales and a constant in the small scales. The reason for this is caused by the fact that the foreground removal techniques tend to remove information from the long-wavelength modes. To calibrate this effect in the large scales, we consider a normalization scale in the power law term, $\ell_\star$. while we just calibrate the transfer function in the small scales by the best fitted constant $C$. If the foreground removal technique is working nicely, this constant should be close to $1$.

We repeat this approach for each map used in this paper. When we run the foreground removal algorithm on a map with foregrounds, we apply this fitted transfer functions $T_{\ell}$ to each measure angular power spectra to correct from the effect of foreground removal on cosmological information.


%

In Fig. \ref{fig:transfer_function} we show the ratio between the cosmological power spectra and the power spectra of the foreground removed maps. We also show the transfer function best fit for the $df=10$ bin configuration.

We show a table with all the best fit values for each frequency bin in Appendix \ref{sec:appendixA}.

\subsubsection{Frequency bin width of $10$ MHz}
We need to understand the physics of the different maps included in this analysis by first measuring the auto-angular power spectra. In Fig. \ref{fig:df10_cls_rec}, we show some examples in order to understand and test our simulated maps. We show the angular power spectra for the $f=700-710$ MHz bin, which corresponds to a redshift range $z=1-1.03$ and it is the highest redshift bin we consider in this bin configuration. In the top left panel of Fig. \ref{fig:df10_cls_rec}, we show the angular power spectra $C_{\ell}$ for the pure cosmological signal (black circles). In red circles, we can see the signal given by the map produced after removing two  components of the fastICA decomposition, while we the signal given by the map after removing three components is given by the blue squares. We can see how we are removing cosmological information, as the amplitude of the power spectrum is smaller. We have considered the highest redshift bin because it is on the edges of our frequency range where the fastICA reconstruction performance is worse.

We can see the effect of the transfer functions in the top right panel. In this case, we have applied the transfer function correction to the angular power spectra of the top left panel and as can be seen, we recover the original information. But this only happens when we run the foreground removal algorithm on the pure cosmology maps. When we include the foreground maps from the GSM model, we do not completely remove the information from the foregrounds, and therefore the amplitude of the angular power spectra of the corrected maps is higher than the one coming only from the large scale structure temperature fluctuations. This can be seen in the right bottom panel. Finally, if we include the noise maps predicted for Tianlai, we can see how the noise signal is the predominant one, even after foreground removal, in the left bottom panel.  The wiggles that can be seen in the bottom left panel in Fig. \ref{fig:df10_cls_rec} are due to the effect of the Tianlai baseline on the measurement beam. This beam presence is seen in any single measurement in maps that include our simulated Tianlai noise maps.


%

\begin{figure}
    \centering
    \begin{tabular}{c}
    \includegraphics[width=0.43\textwidth]{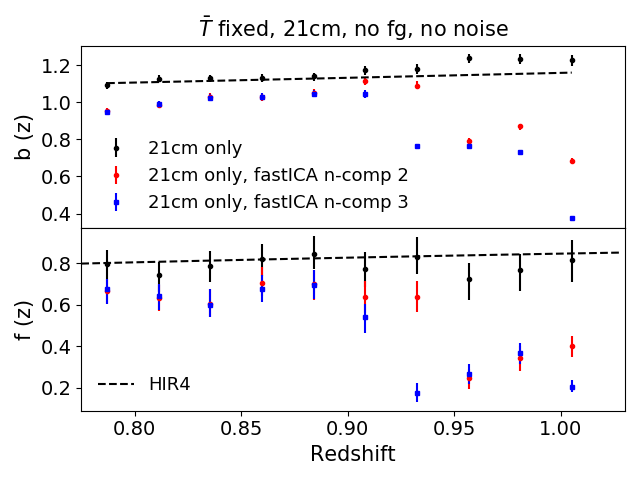}\\
    \includegraphics[width=0.43\textwidth]{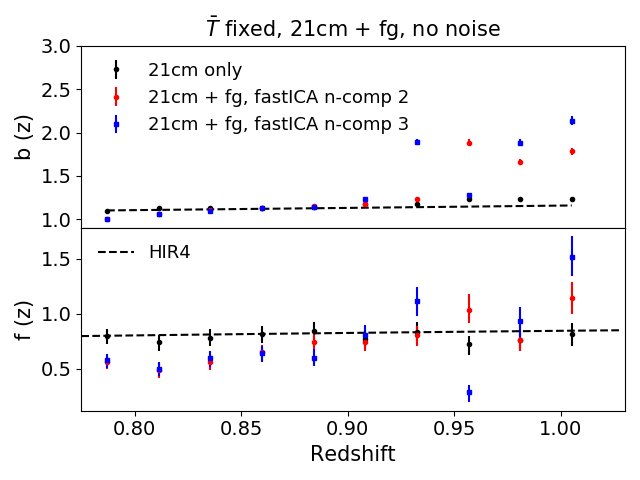}\\
    \includegraphics[width=0.43\textwidth]{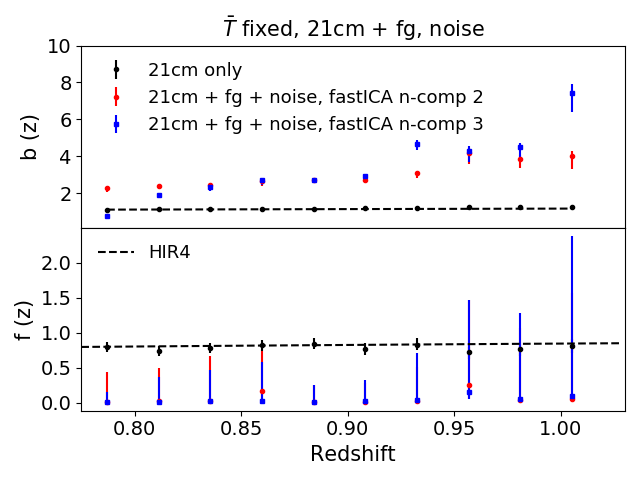}
    \end{tabular}
    \caption{The 68\% confidence limit on the linear bias $b(z)$ and the growth rate of structure $f(z)$ as a function of redshift bin for the $df=10$ MHz case.  Top panel shows the case in which only 21cm pure maps are considered (and no transfer function applied), middle panel considers the case in with foreground signal while the bottom panel shows the constraints when including the noise map. In both middle and lower panel the fit is done by applying a transfer function. We include the HR4 theoretical values (dashed lines).}
    \label{fig:horiz_df10_1}
\end{figure}
\begin{figure}
    \centering
    \begin{tabular}{c}
    \includegraphics[width=0.43\textwidth]{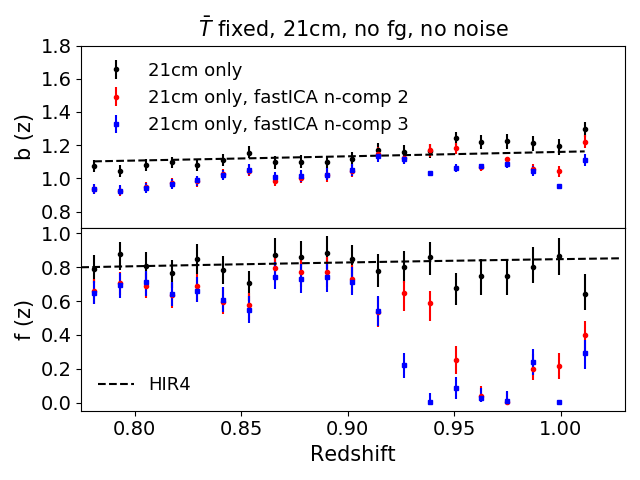}\\
    \includegraphics[width=0.43\textwidth]{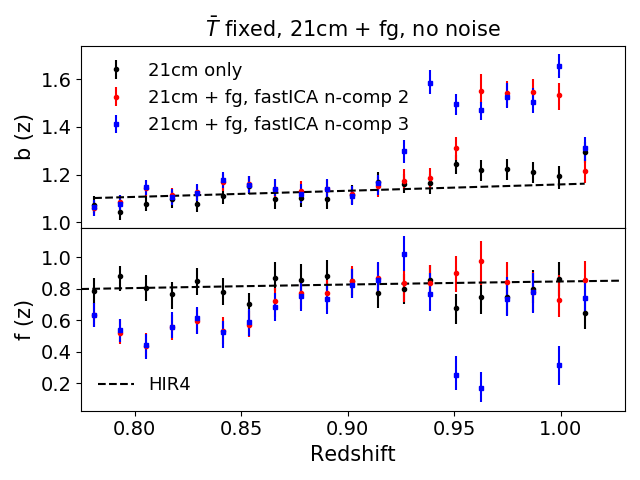}\\
    \includegraphics[width=0.43\textwidth]{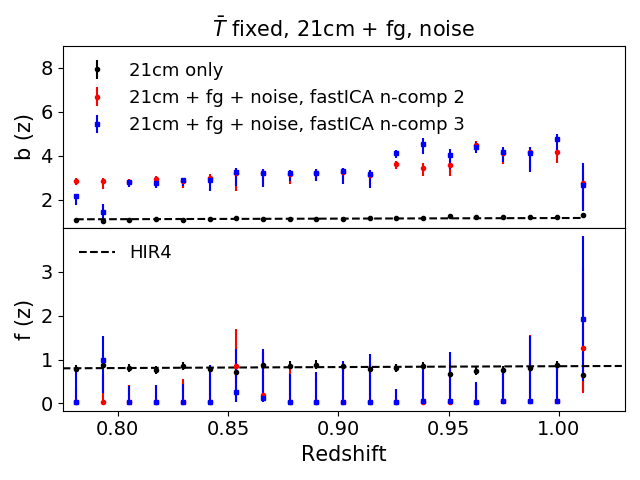}\\
    \end{tabular}
    \caption{The 68\% confidence limit on the linear bias $b(z)$ and the growth rate of structure $f(z)$ as a function of redshift bin for the $df=5$ MHz case. Top panel shows the case with 21cm cosmological maps and two different fastICA reconstructions (without correction). Middle panel considers the case with foreground signal and the bottom panel shows the constraints when including the noise map. We include the simulation input with the dashed lines).}
    \label{fig:horiz_df5_1}
\end{figure}
\begin{figure}
    \centering
    \begin{tabular}{c}
    \includegraphics[width=0.43\textwidth]{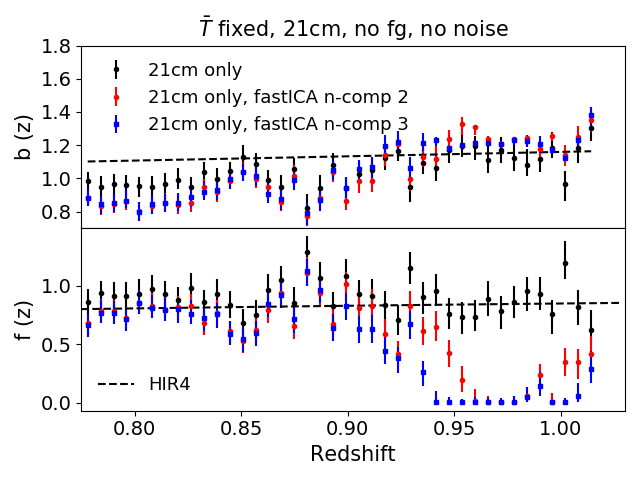}\\
    \includegraphics[width=0.43\textwidth]{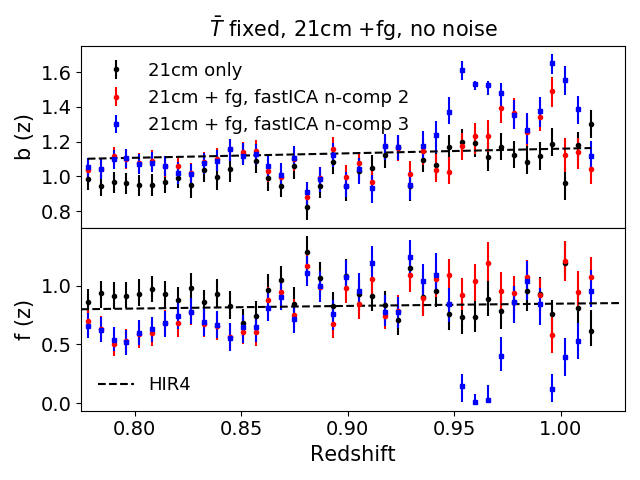}\\
    \includegraphics[width=0.43\textwidth]{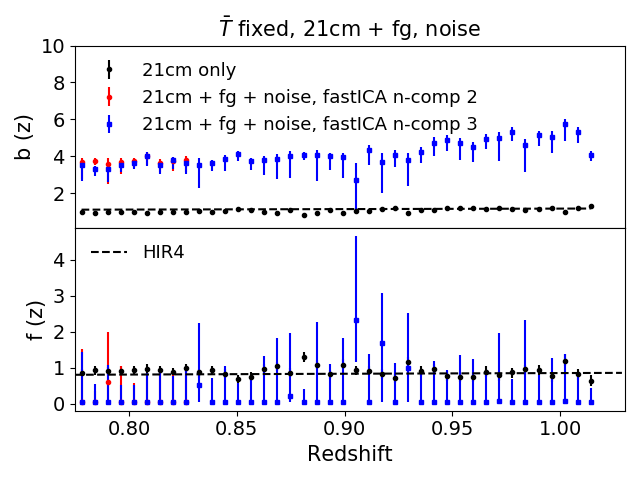}\\
    \end{tabular}
    \caption{Same as Figs. \ref{fig:horiz_df10_1} and  \ref{fig:horiz_df5_1} for the $df=2.5$ MHz case. In the top panel we show the best fit for the case with pure simulated cosmological maps (without corrections). The dashed line represents the simulation input information. Middle panel includes the GSM foreground signal. We include the results for two (red circles) and three (blue squares) fastICA components in the reconstructed maps. The bottom panel shows the best fit when including Tianlai noise.}
    \label{fig:horiz_df2p5_1}
\end{figure}

\begin{figure}
    \centering
    \includegraphics[width=0.43\textwidth]{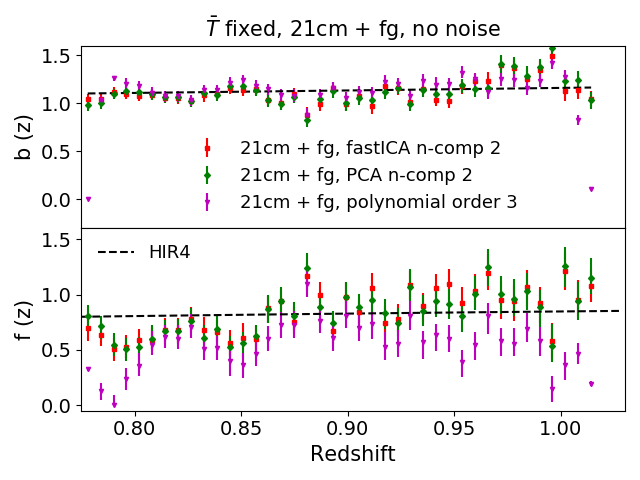}
    \caption{Comparison between foreground removal techniques. We show the constraints on the hydrogen bias and the growth rate for the fastICA, PCA and log-polynomial fitting reconstructed maps, including the transfer functions, for two components and order three for the polynomial.}
    \label{fig:fr_comparison}
\end{figure}


\subsubsection{Frequency bin width of $5$ MHz}
We studied the effect of including more frequency bins in the foreground removal recovery of cosmological information. In Fig. \ref{fig:df5_cls_rec}, we show the angular power spectra for the lowest frequency bin $700-705$ MHz. We can see that even the maps produced directly by the fastICA foreground removal, with no transfer function correction, are closer to the cosmological power spectrum than the maps for the $df=10$ MHz case, as shown in top left panel of Fig. \ref{fig:df5_cls_rec}. This is due to the fact that the bigger the number of frequency bins, the better we trace the smooth evolution with frequency of the foregrounds. We can see in this same panel that when the foreground removal is more efficient, we also remove cosmological information as the amplitude of the angular power spectra of the maps produced by fastICA foreground removal is smaller than the original signal. This is due to the fact that the algorithm is unable to distinguish between the cosmological signal and the radio foregrounds.

By definition, we show in the top right panel that the transfer function allow us to recover the original input. On the bottom left panel, we show the effect of applying foreground removal and transfer functions to the map with foregrounds and we see that we almost recover the original information. Therefore, the increase on the number of frequency bins is significantly improving our reconstruction of the cosmological information encoded in the observed maps.

Finally, we still see that the noise is the main signal on the recovered maps in the bottom right panel, as the amplitude of the noise map is significantly higher than the cosmological signal. This implies that receiver noise and the smoothing caused by the survey strategy is the main systematic regarding the recovery of the original simulation information. 

We also notice that there is almost no difference between both foreground removal maps, either if we remove only two components or three components. If this is the case, then using only the two component maps is a more reasonable option.

\subsubsection{Frequency bin width of $2.5$ MHz}
We finally test the smallest frequency configuration we are considering, the one in which the frequency bins have a bandwidth of $df=2.5$ MHz, which corresponds to $40$ bins in redshift. This is the best sampling of the frequency evolution that we use in this paper. By evaluating Fig. \ref{fig:df2p5_cls_rec}, we see almost no difference with respect to the $df=5$ MHz case shown in Fig. \ref{fig:df5_cls_rec}. The most noticeable difference is the fact that the reconstructed map given by two or three foreground removal components are even more indistinguishable than before. This happens because the more frequency bins we consider, the better the foreground removal technique works and the foregrounds are better removed with less components. Therefore, the conclusions extracted from the previous case are the same here.

\subsection{Cosmological constraints}
In order to understand the cosmological information encoded in the simulation, to benchmark our H1R4 catalogues and to estimate how the foregrounds and the receiver noise affect the constraints on the growth rate of structure. Our procedure consisted on fitting the individual angular power spectrum of each redshift bin, for each bin configuration in the different catalogues introduced in section \ref{sec:mapsobse}. When fitting the cosmological maps given by $T^{\mathrm{obs}}_{b} = T^{\mathrm{HI}}_{b}$ or the maps that include foreground signal, $T^{\mathrm{obs}}_{b} = T^{\mathrm{HI}}_{b}+T^{\mathrm{foreground}}_{b}$ we restrict the fitting of the angular power spectrum to the scales $\ell=20-200$. When we include the noise, the information on the small scales is meaningless for cosmology purposes and we limit our analysis to the scales between $\ell=20-60$. 

In order to include the information from the covariance matrix, we have done a preliminary fit assuming a linear gaussian errors. Then, once we have a best fit values on the bias $b(z)$ and the growth $f(z)$, we created gaussian realizations using a theoretical power spectrum for each bin given by the best fit parameters. When obtaining a most realistic covariance matrix, we have also considered the effect of the mask in which can imply the apparition of off-diagonal elements in the covariance matrix, although they may be small as shown in Fig. \ref{fig:HImapnside128}. The cosmological constraints shown here are the ones that were obtained using the full covariance matrix information.

\subsubsection{Frequency bin width of $10$ MHz}
We have constrained the values of the neutral hydrogen bias and the growth rate of structure for different bin configurations and foreground and noise levels. For each bin configuration, we have increased the layers of complexity by adding foreground signal and receiver noise. In addition, we tested the effect of foreground removal techniques in the cosmological constraints. 

We see on the top panel of Fig. \ref{fig:horiz_df10_1} the constraints on both the hydrogen bias and the growth rate of structure when we only consider the cosmological information in our simulated maps. In this case, we recover the cosmological signal that was used to generate the mock hydrogen catalogues. But we also show what happens if we run the foreground removal algorithm on the 21cm cosmological maps. On the same panel we can see how the best fit values on the bias are smaller than the theoretical expected values. This is due to the fact that with only 10 bins the reconstruction, which is based on segregating the smooth signal from the foreground from the density fluctuations in the temperature field. Let us remind that there is much more fitting bias when measuring the hydrogen bias $b(z)$ than when measuring the growth rate $f(z)$. This happens because the hydrogen bias is mostly constrained by the small scales as it affects all the scales, while the growth rate only affects the larger scales as the linear redshift-space distortions only add a boost in the amplitude at large scales. This is the reason why the bias on the growth rate is much smaller.

When we include the foregrounds and repeat the same fit to the same cosmological parameters, we obtain the plot shown in the middle of Fig. \ref{fig:horiz_df10_1}. In this case, the best fit values for the hydrogen bias continue to be biased with respect to the input theoretical values used to generate the catalogues. Again, the growth rate values are recovered for the same reasons stated above. The only main difference with respect to the previous plot is that we need higher values on the bias in order to fit the observed angular power spectra show in the bottom left panel of Fig. \ref{fig:df10_cls_rec} as the addition of foregrounds in the map makes the foreground removal less efficient.

Finally, in the bottom panel we see that when considering the noise maps, it is impossible to recover the bias information because all the small scale information is destroyed by the beaming. As the bias is not constrained, we obtain best fits on the growth rate consistent with no redshift-space distortions as the noise avoids this possibility.

\subsubsection{Frequency bin width of $5$ MHz}
We checked how the narrowing of the binning affects the reconstruction of the cosmological information after foreground removal. We show in  the top panel of Fig. \ref{fig:horiz_df5_1} how the addition of more frequency bins affects the performance of fastICA. Comparing with the top panel of Fig. \ref{fig:horiz_df10_1}, where there were only ten frequency bins,  we find that the best fit values for the hydrogen bias and growth rate are closer to the input values. This is explained by the better sampling of the evolution of the foreground temperature maps with frequency, allowing for a better reconstruction of the cosmological signal and a more accurate fit.

We also notice a similar pattern in both cases ($df=5$ and $df=10$) in which the constraints for both the bias and the growth rate of structure at higher redshifts are more biased (less accurate) than at lower redshifts. (Compare the top and middle panels of Fig. \ref{fig:horiz_df10_1} and \ref{fig:horiz_df5_1}.)   This happens because the reconstruction method removes more power on all scales at higher redshift than at lower redshift, as can be see on Fig. \ref{fig:transfer_function} for the $df=10$ MHz case. This removed power cannot be easily corrected for, even when using the transfer function, as the reduction in cosmological power is not matched when reconstructing a foreground contaminated sky. In particular, for fastICA, there is the possibility is that  the small number of angular modes present in the sky map on very large-scales means that the power on these scales is not Gaussian distributed (owing to cosmic variance). The fastICA reconstruction method is then flexible enough to `fit'  and remove this non-gaussianity, though it is not clear why this only happens at high redshift.  This problem regarding non-gaussianities is not as large in PCA and log-polynomial fit algorithms, but both of these reconstruction methods also suppress cosmological information on all scales, as they are not able to segregate all foreground components from the cosmological signal and to model completely the foreground signal respectively.

In the middle panel, we show the best fit parameters for the case in which we added the foreground signal to the cosmological maps. Again, the recovery of the growth rate is much better than the recovery of the bias values. But we also notice that we can measure the growth rate of structure with a $10\%$ precision.

When we introduce the noise we do not recover the theoretical value of the bias. As we restrict our fitting at $\ell_{max}=60$, we obtain less offset (or more accurate) fits on the growth rate of structures by degrading the precision of the fit.

\subsubsection{Frequency bin width of $2.5$ MHz}
The configuration with the narrowest bin configuration, $df=2.5$ MHz, should allow for a better reconstruction of the cosmological information as it allows us to sample better the smooth components from the foregrounds. We can see this in the different panels of Fig. \ref{fig:horiz_df2p5_1}. On the top panel we see that for a conservative foreground removal approach we almost do not remove part of the cosmological signal as in the previous cases. This is an improvement with respect to the previous cases due to the better sampling of the frequency range. The reason that removing only two components of the fastICA decomposition works better than for three is due to the fact that when we remove more components the risk of removing cosmological information increases, as it happens in this case.

When we also include the foreground signal in the maps, we also do better foreground removal if only considering two components of the fastICA decomposition. The recovering of the input parameters is also good at higher redshifts as the narrower redshift bin improves the foreground removal. We can measure the growth rate with a precision of $10\%$ again in this case. 

Finally, we see the same pattern than in the $df=5$ case when introducing the noise on the mock maps. By reducing the scales included in the fitting, $\ell = 20-60$, we cannot measure the hydrogen bias as we do not include the small scales but we can measure the growth rate of structures with a $100\%$ precision.

\subsection{Foreground removal comparison}
In the previous results we only considered the maps after foreground removal reconstruction given by fastICA. In this section we explore the alternative reconstruction methods.

In terms of the PCA reconstruction, we found that the results to be very similar to those from fastICA, with the same number of modes. Once again, increasing the number of PCA modes from $n=1$ to $n=2$ had a noticeable impact on the cosmology recovered, significantly reducing the offset between the measured posterior and the posterior for the 21cm only case. Changing from $n=2$ to $n=3$ introduced no significant change in the size of the measured offset, considering all of the bins as a whole.

In terms of the in log-polynomial fitting, we first experimented with various $n$ (maximum order of polynomial), with Eq. (\ref{eq:logpoly}-\ref{eq:polyestimator}). We found that $n=3$ gives the best result in the frequency range {[}700 \textendash{} 800{]} MHz in removing the foreground. This result seems to be due to the smallness of $n$ that is just optimal to mimic the smoothness of the foreground. We also found that the goodness of reconstruction, in terms of the recovered angular power spectrum, depends on the frequency $\nu_{i}$. This reflects the relative weakness in the polynomial fitting, as is also demonstrated by \citet[Fig. 3]{2008MNRAS.388..247D}.  As shown in that paper, the log-polynomial model is not as good as describing the physics of the foregrounds as the log-polynomial model is simpler than the real physical model.

In Fig. \ref{fig:fr_comparison} we show a comparison between the fits for the hydrogen bias, $b$ and the growth rate $f(z)$ when considering the fastICA and the PCA reconstructed maps with $n_{comp}=2$ and when removing components using a polynomial of third order. From the previous results we learned that this number of components is enough in terms of foreground removal and we also consider the case with frequency bins of $df=2.5$ MHz as this configuration produces the best reconstruction. By inspecting the figure, we see that both fastICA and PCA methods provide equally good results while polynomial method does not recover that well the growth rate information, especially on the first and last redshift bins. Therefore, we decided to use fastICA in this study arbitrarily as there is no method that performs better than the other. However, we may consider to use PCA for future analysis as its reliability is also quite good.

\section{Summary}
\label{sec:discussion}

With the rise of radio cosmology, hydrogen intensity mapping has been raised as a promising cosmological probe. Currently there are a number of different SKA pathfinders that are producing the first wide field intensity mapping surveys. In order to understand the systematic errors, which we expect from radio foregrounds and receiver noise and beaming, and the possibilities of improving the cosmological constraints in the near future with this type of survey, we need to develop sophisticated cosmological and observational simulations.

We have created the first simulations of 21cm intensity map signal across the entire sky produced using the HR4 simulations. The simulated catalogues can cover the full sky up to redshift $z=1.5$. To match the frequency range of the Tianlai pathfinder experiment,  in this paper we have focused on the range $700-800$ MHz. Starting from the Friend-of-friends halo catalogue, we have applied a halo model in which the neutral hydrogen mass contained in each halo is given by the dark matter halo mass and the virial velocity of the halo, obtained assuming a spherical collapse model, following the prescription of \cite{2017MNRAS.464.4008P}. In particular, neutral hydrogen populations are suppressed in low-mass halos, as the gas is not bound to the halo, and in more massive halos as the neutral hydrogen gas is heated and becomes excited. Once we have a sample of halos with neutral hydrogen, we convert the mass in a given redshift bin and a angular pixel to a brightness temperature, and generate our set of maps for each frequency band.

To test the consistency of our analysis process, and also to forecast the effectiveness of the maps as a cosmological probe, we measure the angular power spectra and covariance matrix of the 21cm intensity at different redshifts. We use these data products to constrain the hydrogen bias $b_{\rm HI}$ and the growth rate of structure $f$. We show that from the pure 21cm cosmological maps we obtain the same values for these parameters as those predicted assuming the cosmology that generated the original HR4 simulation.

We have also created maps that include the foreground signal as well as the cosmological contribution. The foreground maps are created using the Global Sky Model. This method uses the information from 29 maps at different frequencies and performs a PCA decomposition of 6 components in order to produce foreground maps at any frequency. In particular, for the frequency range we are considering (700-800 MHz), the main foregrounds are synchrotron emission, galactic neutral hydrogen and thermal free-free emission. We have not considered adding any ad hoc information from extragalactic point source emission, part of which should be in the GSM maps.

Once we included the foreground signal, we first masked the galactic center, as the foreground emission is unavoidable here. With the remaining unmasked part, we applied the reconstruction techniques in order to remove the foregrunds and recover the cosmological signal, which were independent component analysis fastICA and principle component analysis PCA. We created recovered maps by removing two or three components.

We  show that when we apply the foreground removal algorithm to the data, we are removing part of the cosmological information, even if we apply it in the case where no foreground is present. Since a strategy is needed to account for the missing power, our chosen option is to define a transfer function that corrects for this. The parameters of the transfer function are fixed by the best fit to the ratio between the original cosmological signal (pure 21cm simulations) and the maps the produced by foreground removal when we apply it directly to the original maps (reconstructed maps). This correction technique becomes more successful as we increase the number of bins, i.e. it works better when the foreground removal is optimal.

We found that in all cases without instrument noise, but where the transfer function correction to the angular power spectrum has been applied, we still recover the input values for the hydrogen bias  growth rate. There is a small degree of offset between the input and recovered values of $b$ and $f$, but this decreases as the number of frequency bins increases, as the foreground reconstruction process becomes more effective for a larger number of bins.

Finally, we considered the effect of noise maps produced for the Tianlai survey. In this case it was impossible to use the small scale part of the angular power spectra for cosmological parameter estimation. When constraining the angular power spectrum, the bias information is set from the amplitude while the information on the growth rate comes from the boost in the low multipole-part of the spectra. If we are unable to recover any cosmological signal on small scales,  this then removes our ability to constrain the hydrogen bias $b(z)$. This in turn diminishes our ability to see any relative change between the large scale and the small scale power due to redshift-space distortions and weakens our constraints on the linear growth rate of structure $f(z)$.

We have shown in this paper when considering the predicted noise present in the Tianlai instrument, we are not able to recover any information on the hydrogen bias, and can only  partially recover the information on the growth of structure through truncating to the large scale information. Enhanced noise removal techniques should be considered in the future in order to fully recover the cosmological information in an unbiased manner.

The presence of noise and foreground residuals can also be mitigated by cross-correlation of the radio intensity map with some optical galaxy catalogue. We will extend this work to use the HR4 simulation to generate a galaxy redshift survey over the same region of sky and redshift, and demonstrate the utility of cross-correlation in accurately recovering the input cosmological parameters \citep{Feng2020}.

\section*{Acknowledgements}
We like to thank Sungwook Hong, Benjamin L'Huillier and Changbom Park for multiple discussions and for providing the HR4 simulations. We thank Glen Rees, Syed Faisal ur Rahman and Song Chen for comments on early software development relating to this project. J. Asorey, D. Parkinson, F. Shi and Y-S. Song are supported by the project \begin{CJK}{UTF8}{mj}우주거대구조를 이용한 암흑우주 연구\end{CJK} (``Understanding Dark Universe Using Large Scale Structure of the Universe''), funded by the Ministry of Science. K. Ahn is supported by NRF-2016R1D1A1B04935414 and NRF-2016R1A5A1013277. L. Zhang is supported by the National Science Foundation of China (11621303, 11653003, 11773021,11890691), the National Key R\&D Program of China (2018YFA0404601, 2018YFA0404504), the 111 project, and the CAS Interdisciplinary Innovation Team (JCTD- 2019-05). This work has been produced using CosKASI Seondeok cluster and the KISTI Nurion machine with project ``HIR4: simulated neutral hydrogen intensity maps from Horizon Run 4 for Future Galaxy Surveys''. 

This research made use of Astropy,\footnote{http://www.astropy.org} a community-developed core Python package for Astronomy \citep{astropy:2013, astropy:2018}, the HEALPix and Healpy package \citep{2005ApJ...622..759G,Zonca2019}, the  Numpy package \cite{book}, the Scipy package \citep{2019arXiv190710121V} and Matplotlib package \citep{Hunter:2007}.

\bibliographystyle{mnras}
\bibliography{references}

\appendix
\section{Transfer function fitting}
\label{sec:appendixA}

\begin{table*}
\centering
\begin{tabular}{ |c|c|c|c|c|c|c|c|c|c|c|c|c| } 
\hline
&\multicolumn{2}{c}{$n_{c}=2$}&\multicolumn{2}{c}{$n_{c}=3$} &\multicolumn{2}{c}{$n_{c}=2$} &\multicolumn{2}{c}{$n_{c}=3$} &\multicolumn{2}{c}{$n_{c}=2$}&\multicolumn{2}{c}{$n_{c}=3$}\\
Freq. (MHz)  & $\ell_{\star}$& $C$& $\ell_{\star}$& C & $\ell_{\star}$& $C$& $\ell_{\star}$& $C$& $\ell_{\star}$& $C$& $\ell_{\star}$& $C$\\
\hline
797.5-800 & \multirow{4}{*}{1.196} &  \multirow{4}{*}{1.327}&\multirow{4}{*}{1.193} &\multirow{4}{*}{1.327} & \multirow{2}{*}{1.104}&\multirow{2}{*}{1.338} &\multirow{2}{*}{1.101} &\multirow{2}{*}{1.338} &0.821 &1.34 &0.82 &1.34\\
  \cline{1-1}\cline{10-13}
  795-797.5 &  &   & & & & & & &1.237 &1.33 &1.228 &1.33\\
  \cline{1-1}\cline{6-13}
  792.5-795  &   & & & &\multirow{2}{*}{0.937} &\multirow{2}{*}{1.317} &\multirow{2}{*}{0.942} &\multirow{2}{*}{1.318} &1.136 &1.327 &1.137 &1.327\\
  \cline{1-1}\cline{10-13}
  790-792.5  &   &   & & & & & & &1.048 &1.326 &1.045 &1.326\\
 \hline
 787.5-790 & \multirow{4}{*}{0.969} &  \multirow{4}{*}{1.285}&\multirow{4}{*}{0.968} &\multirow{4}{*}{1.287} & \multirow{2}{*}{1.564}&\multirow{2}{*}{1.309} &\multirow{2}{*}{1.550} &\multirow{2}{*}{1.309} &2.111 &1.32 &2.079 &1.32\\
  \cline{1-1}\cline{10-13}
  785-787.5 &  &   & & & & & & &0.996 &1.297 &0.954 &1.297\\
  \cline{1-1}\cline{6-13}
  782.5-785  &   & & & &\multirow{2}{*}{1.101} &\multirow{2}{*}{1.3} &\multirow{2}{*}{1.086} &\multirow{2}{*}{1.3} & 1.179&1.304 &1.135 &1.304\\
  \cline{1-1}\cline{10-13}
  780-782.5  &   &   & & & & & & &1.009 &1.297 &0.949 &1.296\\
 \hline
  777.5-780 & \multirow{4}{*}{0.734} &  \multirow{4}{*}{1.223}& \multirow{4}{*}{0.643}&\multirow{4}{*}{1.218} &\multirow{2}{*}{0.87} &\multirow{2}{*}{1.263} &\multirow{2}{*}{0.865} &\multirow{2}{*}{1.263} &0.989 &1.277 &0.989 &1.277\\
  \cline{1-1}\cline{10-13}
  775-777.5 &  &   & & & & & & &0.887 &1.267 &0.877 &1.267\\
  \cline{1-1}\cline{6-13}
  772.5-775  &   & & & &\multirow{2}{*}{0.81} &\multirow{2}{*}{1.241} &\multirow{2}{*}{0.765} &\multirow{2}{*}{1.238} &8.666 &1.26 &7.953 &1.261\\
  \cline{1-1}\cline{10-13}
  770-772.5  &   &   & & & & & & &0.654 &1.231 &0.643 &1.229\\
 \hline
 767.5-770 & \multirow{4}{*}{0.892} &  \multirow{4}{*}{1.215}&\multirow{4}{*}{0.817} &\multirow{4}{*}{1.215} &\multirow{2}{*}{0.814} &\multirow{2}{*}{1.235} &\multirow{2}{*}{0.791} &\multirow{2}{*}{1.234} &0.776 &1.245 & 0.776&1.245\\
  \cline{1-1}\cline{10-13}
  765-767.5 &  &   & & & & & & &0.99 &1.243 &1.001 &1.245\\
  \cline{1-1}\cline{6-13}
  762.5-765  &   & & & &\multirow{2}{*}{1.168} &\multirow{2}{*}{1.219} &\multirow{2}{*}{1.1} &\multirow{2}{*}{1.22} &1.096 &1.221 &1.071 &1.223\\
  \cline{1-1}\cline{10-13}
  760-762.5  &   &   & & & & & & &1.098 &1.216 &1.083 &1.217\\
 \hline
  757.5-760 & \multirow{4}{*}{0.908} &  \multirow{4}{*}{1.2}& \multirow{4}{*}{0.948}&\multirow{4}{*}{1.203} &\multirow{2}{*}{1.024} &\multirow{2}{*}{1.203} &\multirow{2}{*}{0.763} &\multirow{2}{*}{1.198} &0.87 &1.199 &0.857 &1.199\\
  \cline{1-1}\cline{10-13}
  755-757.5 &  &   & & & & & & & 7.189&1.201 &7.367 &1.202\\
  \cline{1-1}\cline{6-13}
  752.5-755  &   & & & &\multirow{2}{*}{1.024} &\multirow{2}{*}{1.182} &\multirow{2}{*}{0.762} &\multirow{2}{*}{1.181} &1.096 &1.185 &0.87 &1.181\\
  \cline{1-1}\cline{10-13}
  750-752.5  &   &   & & & & & & &1.129 &1.169 &0.749 &1.163\\
 \hline
  747.5-750 & \multirow{4}{*}{0.671} &  \multirow{4}{*}{1.109}&\multirow{4}{*}{0.368} &\multirow{4}{*}{1.164} &\multirow{2}{*}{0.834} &\multirow{2}{*}{1.156} &\multirow{2}{*}{0.811} &\multirow{2}{*}{1.157} &2.695 & 1.161&0.719 &1.164\\
  \cline{1-1}\cline{10-13}
  745-747.5 &  &   & & & & & & &1.027 &1.149 &0.555 &1.141\\
  \cline{1-1}\cline{6-13}
  742.5-745  &   & & & &\multirow{2}{*}{0.546} &\multirow{2}{*}{1.101} &\multirow{2}{*}{0.495} &\multirow{2}{*}{1.109} &0.781 &1.123 &0.385 &1.053\\
  \cline{1-1}\cline{10-13}
  740-742.5  &   &   & & & & & & &0.696 &1.114 &0.561 &1.114\\
 \hline
 737.5-740 & \multirow{4}{*}{0.587} &  \multirow{4}{*}{1.162}&\multirow{4}{*}{0.223} &\multirow{4}{*}{2.552} &\multirow{2}{*}{0.679} &\multirow{2}{*}{1.117} &\multirow{2}{*}{0.218} &\multirow{2}{*}{1.137} &0.529 &1.094 &0.534 &1.097\\
  \cline{1-1}\cline{10-13}
  735-737.5 &  &   & & & & & & &0.526 &1.088 &0.152 &0.728\\
  \cline{1-1}\cline{6-13}
  732.5-735  &   & & & &\multirow{2}{*}{0.523} &\multirow{2}{*}{1.079} &\multirow{2}{*}{0.215} &\multirow{2}{*}{1.618} &0.518 &1.08 &0.215 &0.99\\
  \cline{1-1}\cline{10-13}
  730-732.5  &   &   & & & & & & &0.477 &1.049 &0.216 &1.153\\
 \hline
  737.5-730 & \multirow{4}{*}{0.209} &  \multirow{4}{*}{2.311}&\multirow{4}{*}{0.220} &\multirow{4}{*}{2.464} &\multirow{2}{*}{0.207} &\multirow{2}{*}{1.005} &\multirow{2}{*}{0.212} &\multirow{2}{*}{1.445} &0.36 &0.989 &0.213 &1.264\\
  \cline{1-1}\cline{10-13}
  725-727.5 &  &   & & & & & & &0.192 &0.828 &0.211 &1.228\\
  \cline{1-1}\cline{6-13}
  722.5-725  &   & & & &\multirow{2}{*}{0.216} &\multirow{2}{*}{1.483} &\multirow{2}{*}{0.216} &\multirow{2}{*}{1.488} &0.211 & 0.995&0.213 &1.199\\
  \cline{1-1}\cline{10-13}
  720-722.5  &   &   & & & & & & &0.217 &1.163 &0.217 &1.213\\
 \hline
 717.5-720 & \multirow{4}{*}{0.222} &  \multirow{4}{*}{1.853}&\multirow{4}{*}{0.221} &\multirow{4}{*}{2.629} &\multirow{2}{*}{0.218} & \multirow{2}{*}{1.385}&\multirow{2}{*}{0.218} &\multirow{2}{*}{1.481} &0.214 &1.215 &0.214 &1.215\\
  \cline{1-1}\cline{10-13}
  715-717.5 &  &   & & & & & & &0.215 &1.166 &0.215 &1.184\\
  \cline{1-1}\cline{6-13}
  712.5-715  &   & & & &\multirow{2}{*}{0.209} &\multirow{2}{*}{1.367} &\multirow{2}{*}{0.209} &\multirow{2}{*}{1.367} &0.213 &1.135 &0.213 &1.139\\
  \cline{1-1}\cline{10-13}
  710-712.5  &   &   & & & & & & &0.209 &1.106 &0.208 &1.122\\
 \hline
 707.5-710 & \multirow{4}{*}{0.578} &  \multirow{4}{*}{3.196}&\multirow{4}{*}{0.292} &\multirow{4}{*}{10.827} &\multirow{2}{*}{0.214} &\multirow{2}{*}{1.435} &\multirow{2}{*}{0.214} &\multirow{2}{*}{1.979} &0.212 &1.107 &0.213 &1.283\\
  \cline{1-1}\cline{10-13}
  705-707.5 &  &   & & & & & & &0.210 &1.072 &0.213 &1.489\\
  \cline{1-1}\cline{6-13}
  702.5-705  &   & & & &\multirow{2}{*}{0.453} &\multirow{2}{*}{1.153} &\multirow{2}{*}{0.231} &\multirow{2}{*}{1.264} &0.214 &0.904 &0.216 &1.183\\
  \cline{1-1}\cline{10-13}
  700-702.5  &   &   & & & & & & &0.576 &1.015 &0.292 &0.908\\
 \hline

\hline
\end{tabular}

\caption{Best fit parameters of the transfer function for the different bin configuration and the different number of components of the fastICA decomposition used to reconstruct the hydrogen information.}
\label{tab:transfer_function_bf}
\end{table*}

As described in section \ref{sec:transfer_function}, we show in Tab. \ref{tab:transfer_function_bf} the best fit values for the transfer function parameters $\ell_\star$ and $C$ for the different bin configurations and fastICA $n_{comp}=2$ and $n_{comp}=3$.

\bsp	
\label{lastpage}
\end{document}